\documentclass[twocolumns,10pt,final]{IEEEtran}
\usepackage{latexsym}
\usepackage{cite}
\usepackage{amssymb}
\usepackage{bbm}
\usepackage{bm}
\usepackage[usenames,dvipsnames]{color}
\usepackage{amsmath, amssymb}
\usepackage{dsfont}
\usepackage{comment}
\usepackage[dvips]{graphics}
\DeclareMathOperator{\supp}{supp}
\usepackage{graphicx}
\usepackage{lipsum}
\usepackage{psfrag}
\usepackage{units}
\usepackage{setspace}
\usepackage{theorem}
\usepackage{float}
\usepackage{mathtools}
 \allowdisplaybreaks

\newtheorem{definition}{Definition}

\newtheorem{theorem}{Theorem}
\newtheorem{lemma}{Lemma}
\newtheorem{remark}{Remark}

\newtheorem{proposition}{Proposition}

\newcommand\numberthis{\addtocounter{equation}{1}\tag{\theequation}}
{ 
\bibliographystyle{unsrt}
\bibliographystyle{IEEEtran}
\begin{document}

\makeatletter
\newcommand{\vasti}{\bBigg@{3}}
\newcommand{\vast}{\bBigg@{4}}
\newcommand{\Vast}{\bBigg@{5}}
\makeatother
\newcommand{\be}{\begin{equation}}
\newcommand{\ee}{\end{equation}}
\newcommand{\ba}{\begin{align}}
\newcommand{\ea}{\end{align}}
\newcommand{\baa}{\begin{align*}}
\newcommand{\eaa}{\end{align*}}
\newcommand{\bea}{\begin{eqnarray}}
\newcommand{\eea}{\end{eqnarray}}
\newcommand{\beaa}{\begin{eqnarray*}}
\newcommand{\eeaa}{\end{eqnarray*}}
\newcommand{\p}[1]{\left(#1\right)}
\newcommand{\pp}[1]{\left[#1\right]}
\newcommand{\ppp}[1]{\left\{#1\right\}}
\newcommand{\ber}{$\ \mbox{Ber}$}

\title{Semantic-Security Capacity for Wiretap Channels of Type II}

\author{Ziv Goldfeld, \emph{Student Member, IEEE}, Paul Cuff, \emph{Member, IEEE}, and Haim H. Permuter, \emph{Senior Member, IEEE}
	\thanks{
		The work of 
		Z. Goldfeld and H. H. Permuter was supported by an ERC starting grant and the Cyber Security Research Center (CSRC) at Ben-Gurion University of the Negev. The work of P. Cuff was supported by by the National Science Foundation (grant CCF-1350595) and the Air Force Office of Scientific Research (grant FA9550-15-1-0180).
		\newline This paper was presented in part at the 2016 IEEE International Symposium on Information Theory, Barcelona, Spain, and in part 2016 IEEE CS International Conference on Software Science, Technology and Engineering, Beer-Sheva, Israel.
		\newline Z. Goldfeld and H. H. Permuter are with the Department of Electrical and Computer Engineering, Ben-Gurion University of the Negev, Beer-Sheva, Israel (gziv@post.bgu.ac.il, haimp@bgu.ac.il). Paul Cuff is with the Department of Electrical Engineering, Princeton University,
		Princeton, NJ 08544 USA (e-mail: cuff@princeton.edu).
		}}
\maketitle


\begin{abstract}
The secrecy capacity of the type II wiretap channel (WTC II) with a noisy main channel is currently an open problem. Herein its secrecy-capacity is derived and shown to be equal to its semantic-security (SS) capacity. In this setting, the legitimate users communicate via a discrete-memoryless (DM) channel in the presence of an eavesdropper that has perfect access to a subset of its choosing of the transmitted symbols, constrained to a fixed fraction of the blocklength. The secrecy criterion is achieved simultaneously for all possible eavesdropper subset choices. The SS criterion demands negligible mutual information between the message and the eavesdropper's observations even when maximized over all message distributions.

A key tool for the achievability proof is a novel and stronger version of Wyner's soft covering lemma. Specifically, a random codebook is shown to achieve the soft-covering phenomenon with high probability. The probability of failure is doubly-exponentially small in the blocklength. Since the combined number of messages and subsets grows only exponentially with the blocklength, SS for the WTC II is established by using the union bound and invoking the stronger soft-covering lemma. The direct proof shows that rates up to the weak-secrecy capacity of the classic WTC with a DM erasure channel (EC) to the eavesdropper are achievable. The converse follows by establishing the capacity of this DM wiretap EC as an upper bound for the WTC II. From a broader perspective, the stronger soft-covering lemma constitutes a tool for showing the existence of codebooks that satisfy exponentially many constraints, a beneficial ability for many other applications in information theoretic security.
\end{abstract}



\section{Introduction}\label{SEC:introduction}

\par Information theoretic security has adopted the weak-secrecy and the strong-secrecy metrics as a standard for measuring security. Respectively, weak-secrecy and strong-secrecy refer to the normalized and unnormalized mutual information between the secret message and the channel symbol string observed by the eavesdropper. However, recent work argues that, from a cryptographic point of view, both these metrics are insufficient to provide security of applications \cite{Tessaro_SS_WTC2012,Vardy_Semantic_WTC2012}. Their main drawback lies in the assumption that the message is random and uniformly distributed, as real-life messages are neither (messages may be files, votes or any type of structured data, often with low entropy). Semantic-security (SS) \cite{Goldwasser_Semantic_Security1984,Bellare_Semantic_Security1997} is a cryptographic gold standard that was proposed in \cite{Vardy_Semantic_WTC2012} as an adequate alternative and shown to be equivalent to a vanishing unnormalized mutual information for all message distributions. Adopting SS as our secrecy measure, we establish the SS-capacity of the wiretap channel of type II (WTC II) with a noisy main channel, for which even the secrecy-capacity was an open problem until now. On top of that, the SS-capacity and the strong-secrecy-capacity are shown to coincide.



\par Secret communication over noisy channels dates back to Wyner who introduced the degraded wiretap channel (WTC) and derived its weak-secrecy-capacity \cite{Wyner_Wiretap1975}. Csisz{\'a}r and K{\"o}rner extended Wyner's result to the non-degraded WTC \cite{Csiszar_Korner_BCconfidential1978}, which is henceforth referred to as the WTC I. A special instance of the WTC I is when the eavesdropper's observation is an outcome of a discrete-memoryless (DM) erasure channel (EC), which essentially means that he observes a subset of the transmitted symbols which is chosen at random by nature. The WTC II was proposed by Ozarow and Wyner \cite{Wyner_WTCII1984} as a generalization of this instance, where a more powerful eavesdropper selects which subset to observe and security must hold versus all possible subset choices. Thus, the main challenge in establishing security for the WTC II boils down to finding a single sequence of codes that work well for each of the exponentially many subsets the eavesdropper may choose. In \cite{Wyner_WTCII1984}, the authors overcome this difficulty when the main channel is \emph{noiseless} by relying on a unique randomized coset coding scheme in the proof of achievability. The derived rate-equivocation region was also shown to be tight, which solved the noiseless main channel scenario. The WTC II with a general (i.e., possibly \emph{noisy}) DM main channel, however, remained an open problem ever since.

\par A recent endeavor at the optimal secrecy rate of the WTC II with a noisy main channel was presented in \cite{Yener_WTCII2015} (see also \cite{Mihaljevic_WTCII1994,Luo_WTCII2005,Liu_Nested_WTCII2007 ,Calderbank_active_WTCII2009} for related work). Requiring a vanishing \emph{average} error probability and security with respect to the \emph{weak-secrecy} metric (namely, while assuming a uniformly distributed message and a normalized mutual information), the authors of \cite{Yener_WTCII2015} extended the coset coding scheme from \cite{Wyner_WTCII1984} to obtain an inner bound on the rate-equivocation region. An outer bound was also established by assuming that the subset the eavesdropper chooses to observe is revealed to all parties (i.e., to the legitimate users). Specializing these bounds to the maximal equivocation results in an inner and an outer bound on the weak-secrecy-capacity of a general WTC II; these bounds do not match.

\par In this work, we strengthen both the reliability and the security criteria, and derive the \emph{SS-capacity} of the WTC II with a noisy main channel under a vanishing \emph{maximal} error probability requirement. In the heart of the proof stands a stronger version of the soft-covering lemma which is key for the security analysis. Wyner's original soft-covering lemma \cite[Theorem 6.3]{Wyner_Common_Information1975} is a valuable tool for achievability proofs of information theoretic security \cite{Song_Cuff_Secrecy2014,Bloch_Resolvability_Secrecy2013,Cuff_Henchman_Secrecy2014,Cuff_Distortion_Secrecy2014}, resolvability \cite{Han_Verdu_Resolvability1993}, channel synthesis \cite{Cuff_Synthesis2013}, and source coding \cite{Cuff_Song_Likelihood2014} (see also references therein). The result herein sharpens the claim of soft-covering by moving away from an expected value analysis. Instead, we show that a random codebook achieves the soft-covering phenomenon with high probability. The probability of failure is doubly-exponentially small in the blocklength, enabling more powerful applications through the union bound. Specifically, the lemma lets one prove the existence of codebooks that satisfy exponentially many secrecy-related constraints, which, in turn, resolves the difficulty in the security analysis for the WTC II.

\par As a simple preliminary application of the stronger soft-covering lemma, we derive the SS-capacity of the DM-WTC I under a maximal error probability requirement. In \cite{Hayashi_SS_BCConfidential2015}, this result was established in terms of source universal coding  based on the expurgation technique (e.g., cf. \cite[Theorem 7.7.1]{Cover_Thomas}) for the broadcast channel with confidential messages \cite{Csiszar_Korner_BCconfidential1978}, which subsumes the WTC I as a special case. Efficient code constructions with polynomial complexity that achieve the SS-capacity under an average error probability constraint were presented in \cite{Vardy_Semantic_WTC2012} for the DM scenario and in \cite{Ling_SS_Gaussian_WTC2014} for the Gaussian case, while \cite{Tyagi_Gaussian_WTC2014} derived the Gaussian SS-capacity under a maximal error probability constraint. Complexity not being in the scope of this work, we focus on the fundamental limits of semantically-secure communication and give an alternative proof of the WTC I SS-capacity based on the stronger soft-covering lemma and classic wiretap codes. Since the number of secret messages is only exponentially large, the double-exponential decay the lemma provides ensures SS with arbitrarily high probability. In other words, even though a codebook that satisfies exponentially many constraints related to soft-covering is required, the union bound yields that such a codebook exists. This code is then amended to be reliable with respect to the maximal error probability by relying on the well-known expurgation technique (e.g., cf. \cite[Theorem 7.7.1]{Cover_Thomas}).

\par Somewhat surprisingly, our optimal code construction for the WTC II is just the same. Here, SS involves a vanishing unnormalized mutual information (between the message and the eavesdropper's observation), when maximized over all message distributions and eavesdropper's subset choices. However, noting that their combined number grows only exponentially with the blocklenght, the stronger soft-covering lemma is still sharp enough to imply that the probability of an insecure random wiretap code is doubly-exponentially small. As for the WTC I, reliability is upgraded to account for maximal error probability using expurgation. The direct proof shows that any rate up to the weak-secrecy-capacity of the WTC I with a DM-EC\footnote{the erasure probability corresponds to the portion of symbols the eavesdropper in the WTC II does not intercept} to the eavesdropper, is achievable. The converse follows by showing that the weak-secrecy-capacity of this WTC I upper bounds the SS-capacity of the WTC II. An important consequence of the WTC II SS-capacity proof is that Wyner's wiretap codes for the erasure WTC I, are optimal. The binary version of these codes is, in fact, one of the few examples for which there are explicit constructions of practical secure encoders and decoders with optimal performance \cite{Hamming_WTcodes1991,Chalderbank_LDPC_WTcodes2007}.

\par This paper is organized as follows. Section \ref{SEC:preliminaries} provides definitions and basic properties. In Section \ref{SEC:soft_covering} we state the stronger soft-covering lemma and provide its proof. Section \ref{SEC:wiretapI} describes the WTC I and gives an alternative stronger soft-covering lemma based derivation of its SS-capacity. In Section \ref{SEC:wiretapII} we define the WTC II, state its SS-capacity and prove the result. Finally, Section \ref{SEC:summary} summarizes the main achievements and insights of this work.



\section{Notations and Preliminaries}\label{SEC:preliminaries}

\par We use the following notations. Given two real numbers $a,b$, we denote by $[a\mspace{-3mu}:\mspace{-3mu}b]$ the set of integers $\big\{n\in\mathbb{N}\big| \lceil a\rceil\leq n \leq\lfloor b \rfloor\big\}$. We define $\mathbb{R}_+=\{x\in\mathbb{R}|x\geq 0\}$. Calligraphic letters denote sets, e.g., $\mathcal{X}$, the complement of $\mathcal{X}$ is denoted by $\mathcal{X}^c$, while $|\mathcal{X}|$ stands for its cardinality. $\mathcal{X}^n$ denoted the $n$-fold Cartesian product of $\mathcal{X}$. An element of $\mathcal{X}^n$ is denoted by $x^n=(x_1,x_2,\ldots,x_n)$; whenever the dimension $n$ is clear from the context, vectors (or sequences) are denoted by boldface letters, e.g., $\mathbf{x}$. For any $\mathcal{S}\subseteq[1:n]$, we use $\mathbf{x}^\mathcal{S}=(x_i)_{i\in\mathcal{S}}$ to denote the substring of $x^n$ defined by $\mathcal{S}$, with respect to the natural ordering of $\mathcal{S}$. For instance, if $\mathcal{S}=[i:j]$, where $1\leq i< j\leq n$, then $\mathbf{x}^\mathcal{S}=(x_i,x_{i+1},\ldots,x_j)$. 

Let $\big(\Omega,\mathcal{F},\mathbb{P}\big)$ be a probability space, where $\Omega$ is the sample space, $\mathcal{F}$ is the $\sigma$-algebra and $\mathbb{P}$ is the probability measure. Random variables over $\big(\Omega,\mathcal{F},\mathbb{P}\big)$ are denoted by uppercase letters, e.g., $X$, with similar conventions for random vectors. The probability of an event $\mathcal{A}\in\mathcal{F}$ is denoted by $\mathbb{P}(\mathcal{A})$, while $\mathbb{P}(\mathcal{A}\big|\mathcal{B}\mspace{2mu})$ denotes conditional probability of $\mathcal{A}$ given $\mathcal{B}$. We use $\mathds{1}_\mathcal{A}$ to denote the indicator function of $\mathcal{A}$. The set of all probability mass functions (PMFs) on a finite set $\mathcal{X}$ is denoted by $\mathcal{P}(\mathcal{X})$. PMFs are denoted by the capital letter $P$, with a subscript that identifies the random variable and its possible conditioning. For example, for a discrete probability space $\big(\Omega,\mathcal{F},\mathbb{P}\big)$ and two correlated random variables $X$ and $Y$ over that space, we use $P_X$, $P_{X,Y}$ and $P_{X|Y}$ to denote, respectively, the marginal PMF of $X$, the joint PMF of $(X,Y)$ and the conditional PMF of $X$ given $Y$. In particular, $P_{X|Y}$ represents the stochastic matrix whose elements are given by $P_{X|Y}(x|y)=\mathbb{P}\big(X=x|Y=y\big)$. We omit subscripts if the arguments of the PMF are lowercase versions of the random variables. The support of a PMF $P$ and the expectation of a random variable $X$ are denoted by $\supp(P)$ and $\mathbb{E}\big[X\big]$, respectively.

For a discrete measurable space $(\Omega,\mathcal{F})$, a PMF $Q\in\mathcal{P}(\Omega)$ gives rise to a probability measure on $(\Omega,\mathcal{F})$, which we denote by $\mathbb{P}_Q$; accordingly, $\mathbb{P}_Q\big(\mathcal{A})=\sum_{\omega\in\mathcal{A}}Q(\omega)$, for every $\mathcal{A}\in\mathcal{F}$. We use $\mathbb{E}_Q$ to denote an expectation taken with respect to $\mathbb{P}_Q$. For a random variable $X$, we sometimes write $\mathbb{E}_X$ to emphasize that the expectation is taken with respect to $P_X$. For a sequence of random variable $X^n$, if the entries of $X^n$ are drawn in an independent and identically distributed (i.i.d.) manner according to $P_X$, then for every $\mathbf{x}\in\mathcal{X}^n$ we have $P_{X^n}(\mathbf{x})=\prod_{i=1}^nP_X(x_i)$ and we write $P_{X^n}(\mathbf{x})=P_X^n(\mathbf{x})$. Similarly, if for every $(\mathbf{x},\mathbf{y})\in\mathcal{X}^n\times\mathcal{Y}^n$ we have $P_{Y^n|X^n}(\mathbf{y}|\mathbf{x})=\prod_{i=1}^nP_{Y|X}(y_i|x_i)$, then we write $P_{Y^n|X^n}(\mathbf{y}|\mathbf{x})=P_{Y|X}^n(\mathbf{y}|\mathbf{x})$. We often use $Q_X^n$ or $Q_{Y|X}^n$ when referring to an i.i.d. sequence of random variables. The conditional product PMF $Q_{Y|X}^n$ given a specific sequence $\mathbf{x}\in\mathcal{X}^n$ is denoted by $Q_{Y|X=\mathbf{x}}^n$.

The empirical PMF $\nu_{\mathbf{x}}$ of a sequence $\mathbf{x}\in\mathcal{X}^n$ is
\begin{equation}
\nu_{\mathbf{x}}(x)\triangleq\frac{N(x|\mathbf{x})}{n},
\end{equation}
where $N(x|\mathbf{x})=\sum_{i=1}^n\mathds{1}_{\{x_i=x\}}$. We use $\mathcal{T}_\epsilon^{n}(P_X)$ to denote the set of letter-typical sequences of length $n$ with respect to the PMF $P_X$ and the non-negative number $\epsilon$ \cite[Chapter 3]{Massey_Applied}, i.e., we have
\begin{equation}
\mathcal{T}_\epsilon^{n}(P_X)=\Big\{\mathbf{x}\in\mathcal{X}^n\Big|\mspace{5mu}\big|\nu_{\mathbf{x}}(x)-P_X(x)\big|\leq\epsilon P_X(x),\ \forall x\in\mathcal{X}\Big\}.
\end{equation}

The relative entropy between two probability measures $P$ and $Q$ on the same $\sigma$-algebra $\mathcal{F}$ of subsets of the sample space $\mathcal{X}$, with $P\ll Q$ (i.e., $P$ is absolutely continuous with respect to $Q$) is
\begin{equation}
D(P||Q)=\int_\mathcal{X} dP\log\left(\frac{dP}{dQ}\right),\label{EQ:relative_entropy_def}
\end{equation}
       where $\frac{dP}{dQ}$ denotes the Radon-Nikodym derivative between $P$ and $Q$. If the sample space $\mathcal{X}$ is countable, \eqref{EQ:relative_entropy_def} reduces to
\begin{equation}
D(P||Q)=\sum_{x\in\supp(P)}P(x)\log\left(\frac{P(x)}{Q(x)}\right)\label{EQ:relative_entropy_def_discrete}.
\end{equation}

\section{The Stronger Soft-Covering Lemma}\label{SEC:soft_covering}

%

\begin{figure}[t!]
    \begin{center}
        \begin{psfrags}
            \psfragscanon
            \psfrag{A}[][][1]{$W$}
            \psfrag{B}[][][0.95]{\ \ \ \ \ \ \ \ \ \ \ \ $\mathcal{B}_n\mspace{-5mu}=\mspace{-5mu}\big\{\mathbf{u}(w)\big\}$}
            \psfrag{C}[][][1]{\ \ \ \ \ \ \ \ \ $\mathbf{U}(W)$}
            \psfrag{D}[][][1]{\ \ \ \ \ \ \ \ \ \ \ \ $Q_{V|U}$}
            \psfrag{E}[][][1]{\ \ \ \ \ \ \ \ \ $\mathbf{V}\sim P^{(\mathcal{B}_n)}_{\mathbf{V}}$}
            \includegraphics[scale = .43]{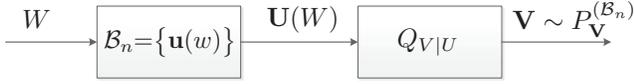}
            \caption{Coding problem with the goal of making $P^{(\mathcal{B}_n)}_{\mathbf{V}}\approx Q_{V}^n$.} \label{FIG:soft_covering}
            \psfragscanoff
        \end{psfrags}
     \end{center}
 \end{figure}


Wyner's soft-covering lemma \cite[Theorem~6.3]{Wyner_Common_Information1975} states that the distribution induced by selecting a $u$-sequence at random from an appropriately chosen set $\mathcal{C}_n$ and passing it through a memoryless channel $Q_{V|U}$, results in a good approximation of $Q_V^n$ in the limit of large $n$, as long as the set is of size $|\mathcal{B}_n|=2^{nR}$, where $R>I(U;V)$ (Fig. \ref{FIG:soft_covering}). In fact, the set can be chosen quite carelessly - by random codebook construction, drawing each sequence independently from the distribution $Q_U^n$.

The soft-covering lemmas in the literature use a distance metric on distributions (commonly total variation or relative entropy) and claim that the distance between the induced distribution $P^{(\mathcal{B}_n)}_{\mathbf{V}}$ and the desired distribution $Q_V^n$ vanishes in expectation over the random selection of the set\footnote{Many of the theorems only claim existence of a good codebook, but all of the proofs use expected value to establish existence.}. In the literature, \cite{Han_Verdu_Resolvability1993} studies the fundamental limits of soft-covering as ``resolvability'', \cite{Hayashi_Secrecy_Resolvability2006} provides rates of exponential convergence, \cite{Cuff_Synthesis2013} improves the exponents and extends the framework, \cite{Ahlswede_Strong_Converse_Quantum2002} and \cite[Chapter 16]{Wilde_Book2013} refer to soft-covering simply as ``covering'' in the quantum context, \cite{Winter_Quantum2005} refers to it as a ``sampling lemma'' and points out that it holds for the stronger metric of relative entropy, and \cite{Kramer_resolvability2013} gives a recent direct proof of the relative entropy result.

Here we give a stronger claim. With high probability with respect to the set construction, the distance vanishes exponentially quickly with the blocklength $n$. The negligible probability of the random set not producing this desired result is doubly-exponentially small.


Let $\mathcal{W}=\big[1:2^{nR}\big]$ and $\mathbb{B}_n = \big\{\mathbf{U}(w)\big\}_{w\in\mathcal{W}}$ be a set of random vectors that are i.i.d. according to $Q_U^n$. We refer to $\mathbb{B}_n$ as the random codebook. Let $\mathcal{C}_n=\big\{\mathbf{u}(w,\mathcal{B}_n)\big\}_{w\in\mathcal{W}}$ denote a realization of $\mathbb{B}_n$. For every fixed $\mathcal{B}_n$, the induced distribution is:
\begin{equation}
    P^{(\mathcal{B}_n)}_{\mathbf{V}}(\mathbf{v})=2^{-nR} \sum_{w\in\mathcal{W}} Q_{V|U}^n\big(\mathbf{v}\big|\mathbf{u}(w,\mathcal{B}_n)\big).
\end{equation}

\begin{lemma}[Stronger Soft-Covering Lemma]\label{LEMMA:soft_covering}
For any $Q_{U}$, $Q_{V|U}$, and $R > I(U;V)$, where $|\mathcal{V}|<\infty$, there exist $\gamma_1,\gamma_2 >0$, such that for $n$ large enough
\begin{equation}
    \mathbb{P}\bigg(D\Big(P^{(\mathbb{B}_n)}_{\mathbf{V}}\Big|\Big|Q_V^n\Big)> e^{-n\gamma_1}\bigg)\leq e^{- e^{n\gamma_2}}.\label{EQ:soft_covering}
\end{equation}
More precisely, for any $n \in \mathbb{N}$ and $\delta \in \big(0,R-I(U;V)\big)$
\begin{equation}
\mathbb{P}\bigg(D\Big(P^{(\mathbb{B}_n)}_{\mathbf{V}}\Big|\Big|Q_V^n\Big)> c_\delta n2^{-n\gamma_{\delta}} \bigg)\leq \big( 1 + |{\cal V}|^n \big) e^{-\frac{1}{3} 2^{n\delta}},\label{EQ:soft_covering_precise}
\end{equation}
where
\begin{subequations}
\begin{align}
    \gamma_{\delta} &= \sup_{\alpha > 1} \frac{\alpha - 1}{2 \alpha - 1} \big(R - \delta - d_{\alpha} (Q_{U,V},Q_U Q_V)\big), \label{EQ:soft_covering_exponent}\\
    c_\delta &= 3\log e+2\gamma_\delta \log2+2 \log \left( \max_{v \in\supp(Q_V)} \frac{1}{Q_V(v)} \right),\label{EQ:soft_covering_coefficient1}
\end{align}
and $d_{\alpha}(\Gamma,\Pi)=\frac{1}{\alpha-1}\log_2\int d\mspace{2mu}\Gamma\left(\frac{d\mspace{2mu}\Pi}{d\mspace{2mu}\Gamma}\right)^{1-\alpha}$ is the R\'{e}nyi divergence of order $\alpha$.
\end{subequations}
\end{lemma}

\begin{remark}
The inequality \eqref{EQ:soft_covering_precise} is trivially true for $\delta$ outside of the expressed range.
\end{remark}

The important quantity in the lemma above is $\gamma_{\delta}$, which is the exponent that soft-covering achieves.  We see in \eqref{EQ:soft_covering_precise} that the double-exponential convergence of probability occurs with exponent $\delta>0$.  Thus, the best soft-covering exponent that the lemma achieves with confidence, over all $\delta>0$, is
\begin{equation}
    \gamma^* = \sup_{\delta>0} \gamma_{\delta} = \gamma_0 = \sup_{\alpha > 1} \frac{\alpha - 1}{2 \alpha - 1} \big(R - d_{\alpha} (Q_{U,V},Q_U Q_V)\big).
\end{equation}
The double-exponential confidence rate $\delta$ acts as a reduction in codebook rate $R$ in the definition of $\gamma_{\delta}$.  Consequently, $\gamma_{\delta}=0$ for $\delta \geq R-I(U;V)$.

\begin{remark}[Total Variation Exponent of Decay]
The stronger soft-covering lemma can be reproduced while replacing the relative divergence with total variation \cite{Cuff_SCL_ISIT2016}. Although, relative entropy can be used to bound total variation via Pinsker's inequality, this approach causes a loss of a factor of 2 in the exponent of decay. Alternatively, the proof of Lemma \ref{LEMMA:soft_covering} can be modified to produce the bound on the total variation instead of the relative entropy. This direct method keeps the error exponents the same for the total variation case as it is for relative entropy.
\end{remark}

Before proving Lemma \ref{LEMMA:soft_covering}, we note that the name `stronger soft-covering lemma' is justified because \eqref{EQ:soft_covering} implies that the expectation of the relative entropy over the ensemble of codebooks decays exponentially fast (i.e., Wyner's notion of soft-covering). This is stated in the following lemma and proven in Appendix \ref{APPEN:soft_covering_stronger_proof}.

\begin{lemma}[Stronger than Wyner's Soft-Covering Lemma]\label{LEMMA:soft_covering_stronger} Let $\gamma_1,\gamma_2>0$ be such that \eqref{EQ:soft_covering} holds for $n$ large enough, then for every such $n$,
\begin{equation}
\mathbb{E}_{\mathbb{B}_n}D\Big(P^{(\mathbb{B}_n)}_{\mathbf{V}}\Big|\Big|Q_V^n\Big)\leq e^{-n\gamma_1}+n\log\left(\frac{1}{\mu_V}\right)e^{-e^{n\gamma_2}},
\end{equation}
where $\mu_v=\min_{v\in\supp(Q_V)}Q_V(v)>0$.
\end{lemma}

\begin{IEEEproof}[Proof of Lemma \ref{LEMMA:soft_covering}]
    We state the proof in terms of arbitrary distributions (not necessarily discrete).  When needed, we will specialize to the case that ${\cal V}$ is finite. For any fixed codebook $\mathcal{C}_n$, let the Radon-Nikodym derivative between the induced and desired distributions be denoted as
    \begin{equation}
        \Delta_{\mathcal{B}_n}(\mathbf{v}) \triangleq \frac{d P^{(\mathcal{B}_n)}_{\mathbf{V}}}{d Q_V^n}(\mathbf{v}).\label{EQ:RD_derivative}
    \end{equation}
    In the discrete case, this is just a ratio of probability mass functions. Accordingly, the relative entropy of interest, which is a function of the codebook $\mathcal{B}_n$, is given by
    \begin{equation}
        D\Big(P^{(\mathcal{B}_n)}_{\mathbf{V}}\Big|\Big|Q_V^n\Big)=\int dP^{(\mathcal{B}_n)}_{\mathbf{V}}\log \Delta_{\mathcal{B}_n}.
    \end{equation}

To describe the jointly-typical set over $u$- and $v$-sequences, we first define information density $i_{Q_{U,V}}$, which is a function on the space $\mathcal{U}\times\mathcal{V}$ specified by
\begin{equation}
  i_{Q_{U,V}}(u,v)\triangleq\log\left(\frac{dQ_{V|U=u}}{dQ_V}(v)\right).\label{EQ:information_density}
\end{equation}
In \eqref{EQ:information_density}, the argument of the logarithm is the Radon-Nikodym derivative between $Q_{V|U=u}$ and $Q_V$. Let $\epsilon\geq0$ be arbitrary, to be determined later, and define
    \begin{equation}
        {\cal A}_{\epsilon} \triangleq \left\{ (\mathbf{u},\mathbf{v})\in\mathcal{U}^n\mspace{-4mu}\times\mspace{-3mu}\mathcal{V}^n\mspace{3mu} \bigg|\mspace{3mu}\frac{1}{n} i_{Q^n_{U,V}}(\mathbf{u},\mathbf{v}) < I(U;V) + \epsilon \right\},\label{EQ:typical_set}
    \end{equation}
and note that
\begin{equation}
  i_{Q^n_{U,V}}(\mathbf{u},\mathbf{v})=\sum_{t=1}^ni_{Q_{U,V}}(u_t,v_t).\label{EQ:information_density_product}
\end{equation}

We split $P^{(\mathcal{B}_n)}_{\mathbf{V}}$ into two parts, making use of the indicator function. For every $\mathbf{v}\in\mathcal{V}^n$, define
\begin{subequations}
    \begin{align}
        P_{\mathcal{B}_n,1}(\mathbf{v})\mspace{-4mu}&\triangleq 2^{-nR}\mspace{-3mu} \sum_{w\in\mathcal{W}} \mspace{-4mu}Q_{V|U}^n\big(\mathbf{v}\big|\mathbf{u}(w,\mathcal{B}_n)\big)\mathds{1}\mspace{-3mu}_{\big\{\mspace{-2mu}\big(\mathbf{u}(w,\mathcal{B}_n),\mathbf{v}\big)\in\mathcal{A}_\epsilon\mspace{-2mu}\big\}}, \\
        P_{\mathcal{B}_n,2}(\mathbf{v})\mspace{-4mu}&\triangleq 2^{-nR} \sum_{w\in\mathcal{W}} \mspace{-4mu}Q_{V|U}^n\big(\mathbf{v}\big|\mathbf{u}(w,\mathcal{B}_n)\big)\mathds{1}\mspace{-3mu}_{\big\{\mspace{-2mu}\big(\mathbf{u}(w,\mathcal{B}_n),\mathbf{v}\big)\notin\mathcal{A}_\epsilon\mspace{-2mu}\big\}}.
    \end{align}
\end{subequations}
    The measures $P_{\mathcal{B}_n, 1}$ and $P_{\mathcal{B}_n, 2}$ on the space $\mathcal{V}^n$ are not probability measures, but $P_{\mathcal{B}_n,1}+P_{\mathcal{B}_n,2}=P^{(\mathcal{B}_n)}_{\mathbf{V}}$ for each codebook ${\mathcal{B}_n}$. We also split $\Delta_{\mathcal{B}_n}$ into two parts. Namely, for every $\mathbf{v}\in\mathcal{V}^n$, we set
\begin{subequations}
\begin{align}
    \Delta_{\mathcal{B}_n,1}(\mathbf{v})&\triangleq\frac{dP_{\mathcal{B}_n,1}}{dQ_V^n}(\mathbf{v})\\
    \Delta_{\mathcal{B}_n,2}(\mathbf{v})&\triangleq\frac{dP_{\mathcal{B}_n,2}}{dQ_V^n}(\mathbf{v}).
\end{align}\label{EQ:delta_def}
\end{subequations}
\indent With respect to the above definitions, Lemma \ref{LEMMA:soft_covering_UB} states an upper bound on the relative entropy of interest.
    \begin{lemma}\label{LEMMA:soft_covering_UB}
    For every fixed codebook $\mathcal{B}_n$, we have
    \begin{align*}
        D\Big(P&^{(\mathcal{B}_n)}_{\mathbf{V}}\Big|\Big|Q_V^n\Big)\leq h\left(\int dP_{\mathcal{B}_n,1}\right)\\
        &+ \int dP_{\mathcal{B}_n,1}\log\Delta_{\mathcal{B}_n,1}+\int dP_{\mathcal{B}_n,2}\log\Delta_{\mathcal{B}_n,2}, \numberthis\label{EQ:expanded divergence bound}
    \end{align*}
    where $h(\cdot)$ is the binary entropy function.
    \end{lemma}
    The proof is relegated to Appendix \ref{APPEN:soft_covering_UB_proof}. Based on Lemma \ref{LEMMA:soft_covering_UB}, if the relative entropy of interest does not decay exponentially fast, then the same is true for the terms on the right-hand side (RHS) of (\ref{EQ:expanded divergence bound}). Therefore, to establish Lemma \ref{LEMMA:soft_covering}, its suffices to show that the probability (with respect to a random codebook) of the RHS not vanishing exponentially fast to 0 as $n\to\infty$, is double-exponentially small.

    Notice that $P_{\mathcal{B}_n, 1}$ usually contains almost all of the probability. That is, for any fixed $\mathcal{B}_n$, we have
    \begin{align*}
	&\int dP_{\mathcal{B}_n, 2}=1-\int dP_{\mathcal{B}_n,1}\\&=\sum_{w\in\mathcal{W}} 2^{-nR} \mathbb{P}_{Q_{V|U}^n}\Big(\big(\mathbf{u}(w,\mathcal{B}_n),\mathbf{V}\big)\notin\mathcal{A}_\epsilon\Big|\mathbf{U}=\mathbf{u}(w,\mathcal{B}_n)\Big).\numberthis\label{EQ:chernoff_pre_rvs1}
    \end{align*}
    For a random codebook, (\ref{EQ:chernoff_pre_rvs1}) becomes
    \begin{align*}
	&\int dP_{\mathbb{B}_n, 2}\\&=\sum_{w\in\mathcal{W}}\mspace{-5mu} 2^{-nR} \mathbb{P}_{Q_{V|U}^n}\Big(\big(\mathbf{U}(w,\mathbb{B}_n),\mathbf{V}\big)\notin\mathcal{A}_\epsilon\Big|\mathbf{U}=\mathbf{U}(w,\mathbb{B}_n)\Big).\numberthis\label{EQ:chernoff_rvs1}
    \end{align*}
The RHS of (\ref{EQ:chernoff_rvs1}) is an average of exponentially many i.i.d. random variables bounded between 0 and 1. Furthermore, the expected value of each one is the exponentially small probability of correlated sequences being atypical:
\begin{align*}
	\mathbb{E}_{\mathbb{B}_n}\mathbb{P}_{Q_{V|U}^n}&\Big(\big(\mathbf{U}(w,\mathbb{B}_n),\mathbf{V}\big)\notin\mathcal{A}_\epsilon\Big|\mathbf{U}=\mathbf{U}(w,\mathbb{B}_n)\Big)\\&=\mathbb{P}_{Q_{U,V}^n}\Big(\big(\mathbf{U},\mathbf{V}\big)\notin\mathcal{A}_\epsilon\Big)\\
    &= \mathbb{P}_{Q_{U,V}^n} \left( \sum_{t=1}^n i_{Q_{U,V}}(U_t,V_t) \geq n \big(I(U;V) + \epsilon\big) \right) \\
    &\stackrel{(a)}= \mathbb{P}_{Q_{U,V}^n} \left( 2^{\lambda\sum_{t=1}^n i_{Q_{U,V}}(U_t,V_t)} \geq 2^{n\lambda(I(U;V) + \epsilon)}\right) \\
    &\stackrel{(b)}\leq \frac{\mathbb{E}_{Q_{U,V}^n} 2^{\lambda \sum_{t=1}^ni_{Q_{U,V}}(U_t,V_t)}}{2^{n\lambda (I(U;V) + \epsilon)}} \\
    &= \left( \frac{\mathbb{E}_{Q_{U,V}} 2^{\lambda i_{Q_{U,V}}(U,V)}}{2^{\lambda (I(U;V) + \epsilon)}} \right)^n \\
    &\stackrel{(c)}= 2^{n \lambda \left( \frac{1}{\lambda} \log_2 \mathbb{E}_{Q_{U,V}} \big[2^{\lambda i_{Q_{U,V}}(U;V)}\big] - I(U;V) - \epsilon \right)} \\
    &\stackrel{(d)}= 2^{n \lambda \big( d_{\lambda+1}(Q_{U,V},Q_U Q_V) - I(U;V) - \epsilon \big)},\numberthis\label{EQ:atypical_expectation_UB}
\end{align*}
where (a) is true for any $\lambda\geq0$, (b) is Markov's inequality, (c) follows by restricting $\lambda$ to be strictly positive, while (d) is from the definition of the R\'{e}nyi divergence of order $\lambda+1$. We use units of bits for mutual information and R\'{e}nyi divergence to coincide with the base two expression of rate. Now, substituting $\alpha = \lambda+1$ into \eqref{EQ:atypical_expectation_UB} gives
\begin{subequations}
\begin{equation}
    \mathbb{E}_{\mathbb{B}_n}\mspace{-1mu}\mathbb{P}_{Q_{V|U}^n}\mspace{-3mu}\Big(\big(\mathbf{U}(w,\mathbb{B}_n),\mathbf{V}\big)\mspace{-3mu}\notin\mspace{-3mu}\mathcal{A}_\epsilon\Big|\mathbf{U}=\mathbf{U}(w,\mathbb{B}_n)\Big)\leq 2^{-n\beta_{\alpha,\epsilon}},\label{EQ:atypical probability expectation}
\end{equation}
where
\begin{equation}
    \beta_{\alpha,\epsilon}=(\alpha - 1) \big( I(U;V) + \epsilon - d_{\alpha}(Q_{U,V}, Q_U Q_V) \big),\label{EQ:chernoff_rvs1_prop_beta}
\end{equation}\label{EQ:chernoff_rvs1_prop}
\end{subequations}
\vspace{-6mm}

\noindent for every $\alpha>1$ and $\epsilon\geq 0$, over which we may optimize. The optimal choice of $\epsilon$ is apparent when all bounds of the proof are considered together (some yet to be derived), but the formula may seem arbitrary at the moment.  Nevertheless, fix $\delta \in \big(0,R-I(U;V)\big)$, as found in the theorem statement, and set
\begin{equation}
    \epsilon_{\alpha,\delta}= \frac{ \frac{1}{2} (R-\delta) + (\alpha - 1) d_{\alpha} (Q_{U,V},Q_U Q_V) }{ \frac{1}{2} + (\alpha- 1) } - I (U;V).\label{EQ:optimized_epsilon}
\end{equation}
Substituting into $\beta_{\alpha,\epsilon}$ gives
\begin{equation}
    \beta_{\alpha,\delta}\triangleq\beta_{\alpha,\epsilon_{\alpha,\delta}}=\frac{\alpha - 1}{2 \alpha - 1} \big(R-\delta - d_{\alpha} (Q_{U,V},Q_U Q_V)\big).
\end{equation}
Observe that $\epsilon_{\alpha,\delta}$ in \eqref{EQ:optimized_epsilon} is nonnegative under the assumption that $R-\delta> I(U;V)$, because $\alpha>1$ and $d_{\alpha} (Q_{U,V},Q_U Q_V) \geq d_{1} (Q_{U,V},Q_U Q_V) = I(U;V)$.

Next, we use the following version of the Chernoff bound to bound the probability of (\ref{EQ:chernoff_rvs1}) not being exponentially small. 

\begin{lemma}[Chernoff Bound]\label{LEMMA:Chernoff}
Let $\big\{X_m\big\}_{m=1}^M$ be a collection of i.i.d. random variables with $X_m\in[0,B]$ and $\mathbb{E}X_m\leq\mu\neq 0$, for all $m\in[1:M]$. Then for any $c$ with $\frac{c}{\mu} \in [1,2]$,
\begin{equation}
    \mathbb{P} \left( \frac{1}{M} \sum_{m=1}^M X_m \geq c \right) \leq e^{-\frac{M \mu}{3 B} \left( \frac{c}{\mu} - 1 \right)^2}.\label{EQ:Chernoff}
\end{equation}
\end{lemma}
The proof is given in Appendix \ref{APPEN:chernoff_proof}.

Using \eqref{EQ:Chernoff} with $M = 2^{nR}$, $\mu = 2^{-n\beta_{\alpha,\delta}}$, $B=1$, and $\frac{c}{\mu} = 2$, assures that $\int d P_{{\mathcal{B}_n}, 2}$ is exponentially small with probability doubly-exponentially close to 1. That is
\begin{equation}
    \mathbb{P} \left(\int dP_{\mathcal{B}_n,2}\geq 2 \cdot 2^{- n\beta_{\alpha,\delta}}  \right)\leq e^{-\frac{1}{3}2^{n(R-\beta_{\alpha,\delta})}}.\label{EQ:atypical probability bound_gamma}
\end{equation}

Similarly, $\Delta_{\mathbb{B}_n,1}$ is an average of exponentially many i.i.d. and uniformly bounded functions, each one determined by one sequence in the random codebook:
\begin{align*}
    \Delta&_{\mathbb{B}_n, 1}(\mathbf{v})\\&\mspace{-19mu}=\sum_{w\in\mathcal{W}}2^{-nR}\frac{d Q_{V|U=\mathbf{U}(w,\mathbb{B}_n)}^n}{d Q_V^n} (\mathbf{v}) \mathbf{1}_{\big\{\big(\mathbf{U}(w,\mathbb{B}_n),\mathbf{v}\big)\in\mathcal{A}_\epsilon\big\}}.\numberthis\label{EQ:chernoff_rvs2}
\end{align*}
For every term in the average, the indicator function bounds the value to be between $0$ and $2^{n(I(U;V)+\epsilon_{\alpha,\delta})}$. The expected value of each term with respect to the codebook is bounded above by one, which is observed by removing the indicator function. Therefore, the Chernoff bound assures that $\Delta_{\mathbb{B}_n,1}$ is exponentially close to one for every $\mathbf{v}\in\mathcal{V}^n$. Setting $M = 2^{nR}$, $\mu = 1$, $B = 2^{n(I(U;V) + \epsilon_{\alpha,\delta})}$, and $\frac{c}{\mu} =1+ 2^{-n\beta_{\alpha,\delta}}$ into \eqref{EQ:Chernoff}, gives
\begin{align*}
    \mathbb{P}\Big( \Delta_{\mathbb{B}_n,1}(\mathbf{v}) \geq 1+2^{-n\beta_{\alpha,\delta}}\Big)&\leq e^{-\frac{1}{3} 2^{n(R-I(U;V)-\epsilon_{\alpha,\delta}-2\beta_{\alpha,\delta})}}\\&=e^{-\frac{1}{3} 2^{n\delta}},\quad \forall\mspace{3mu} \mathbf{v}\in\mathcal{V}^n,\numberthis\label{EQ:typical set bound}
\end{align*}
which decays doubly-exponentially fast for any $\delta >0$.

At this point, we specialize to a finite set $\mathcal{V}$. Consequently, $\Delta_{\mathbb{B}_n,2}$ is bounded as
\begin{equation}
    \Delta_{\mathbb{B}_n,2}(\mathbf{v}) \leq \left( \max_{v \in \supp(Q_V)} \frac{1}{Q_V(v)} \right)^n,\ \forall\mspace{3mu} \mathbf{v}\in\mathcal{V}^n,\label{EQ:atypical divergence bound}
\end{equation}
with probability 1. Notice that the maximum is only over the support of $Q_V$, which makes this bound finite. The underlying reason for this restriction is that with probability one a conditional distribution is absolutely continuous with respect to its associated marginal distribution.

Having \eqref{EQ:atypical probability bound_gamma}, \eqref{EQ:typical set bound} and \eqref{EQ:atypical divergence bound}, we can now bound the probability that the RHS of \eqref{EQ:expanded divergence bound} is not exponentially small. Let $\mathcal{S}$ be the set of codebooks $\mathcal{B}_n$, such that all of the following are true:
\begin{subequations}
    \begin{align}
        \int d P_{\mathcal{B}_n,2}&<2\cdot 2^{- n\beta_{\alpha,\delta}},\label{EQ:codebook_S1}\\
        \Delta_{\mathcal{B}_n,1}(\mathbf{v})&<1+2^{-n\beta_{\alpha,\delta}},\ \forall\mspace{3mu}\mathbf{v}\in\mathcal{V}^n,\label{EQ:codebook_S2}\\
        \Delta_{\mathcal{B}_n,2}(\mathbf{v})&\leq\left(\max_{v\in\supp(Q_V)}\frac{1}{Q_V(v)} \right)^n,\ \forall\mspace{3mu}\mathbf{v}\in\mathcal{V}^n.\label{EQ:codebook_S3}
    \end{align}
\end{subequations}
First, we use the union bound, while taking advantage of the fact that the space $\mathcal{V}^n$ is only exponentially large, to show that the probability of a random codebook not being in $\mathcal{S}$ is double-exponentially small:
\begin{align*}
    \mathbb{P}\big(\mathbb{B}_n &\notin \mathcal{S}\big)\\
    &\stackrel{(a)}\leq\mathbb{P}\bigg(\int dP_{\mathbb{B}_n,2}\geq 2\cdot2^{-n\beta_{\alpha,\delta}}\bigg)\\&\mspace{20mu}+\sum_{\mathbf{v}\in\mathcal{V}^n}\mathbb{P}\bigg(\Delta_{\mathbb{B}_n,1}(\mathbf{v})\geq1+2^{-\beta_{\alpha,\delta}  n}\bigg)\\&\mspace{40mu}+\sum_{\mathbf{v}\in\mathcal{V}^n}\mathbb{P}\Bigg(\Delta_{\mathbb{B}_n,2}(\mathbf{v})>\left(\max_{v\in\supp(Q_V)}\frac{1}{Q_V(v)} \right)^n\Bigg)\\
    &\stackrel{(b)}\leq e^{-\frac{1}{3}2^{n(R-\beta_{\alpha,\delta})}} + |\mathcal{V}|^n\cdot e^{-\frac{1}{3} 2^{n\delta}}\\
    &\stackrel{(c)}\leq \left(1+|\mathcal{V}|^n\right)e^{-\frac{1}{3} 2^{n\delta}},\numberthis\label{EQ:notin_S_probability_UB}
\end{align*}
where (a) is the union bound, (b) uses \eqref{EQ:atypical probability bound_gamma}, \eqref{EQ:typical set bound} and \eqref{EQ:atypical divergence bound}, while (c) follows because $\beta_{\alpha,\delta} \leq \frac{1}{2}(R-\delta)$.

Next, we claim that for every codebook in $\mathcal{S}$, the RHS of (\ref{EQ:expanded divergence bound}) is exponentially small. Let $\mathcal{B}_n\in\mathcal{S}$ and consider the following. For every $x\in[0,1]$, $h(x) \leq x \log \frac{e}{x}$, using which \eqref{EQ:codebook_S1} implies that
\begin{align*}
    h\left( \int dP_{\mathcal{B}_n,1} \right)& = h\left( \int d P_{{\mathcal{B}_n},2} \right)\\& < 2\big(\log e+\beta_{\alpha,\delta}\log2\big)n2^{-n\beta_{\alpha,\delta}}.\numberthis\label{EQ:codebook_S_UB1}
\end{align*}
Furthermore, by \eqref{EQ:codebook_S2}, we have
\begin{align*}
    \int d P_{\mathcal{B}_n,1} \log \Delta_{\mathcal{B}_n,1}& < \int dP_{\mathcal{B}_n,1}\log (1 + 2^{-n\beta_{\alpha,\delta}})\\&\leq \log (1 + 2^{-n\beta_{\alpha,\delta}})\\&\stackrel{(a)}\leq 2^{-n\beta_{\alpha,\delta}} \log e,\numberthis\label{EQ:codebook_S_UB2}
\end{align*}
where (a) follows since $\log(1+x)\leq x\log e$, for every $x>0$. Finally, using \eqref{EQ:codebook_S3} we obtain
\begin{align*}
    \int d P_{{\mathcal{B}_n},2} \log \Delta_{{\mathcal{B}_n},2}&\leq \int d P_{{\mathcal{B}_n},2} \log \left( \max_{v \in\supp(Q_V)} \frac{1}{Q_V(v)} \right)^n\\
    &< 2 \log \left( \max_{v \in\supp(Q_V)} \frac{1}{Q_V(v)} \right) n2^{-n\beta_{\alpha,\delta}}.\numberthis\label{EQ:codebook_S_UB3}
\end{align*}

Combining \eqref{EQ:codebook_S_UB1}-\eqref{EQ:codebook_S_UB3}, yields
\begin{align*}
    &h\left(\int dP_{\mathcal{B}_n,1}\right)\mspace{-3mu}+\mspace{-3mu} \int dP_{\mathcal{B}_n,1}\log\Delta_{\mathcal{B}_n,1}\mspace{-3mu}+\mspace{-3mu}\int dP_{\mathcal{B}_n,2}\log\Delta_{\mathcal{B}_n,2}\\
    &\begin{multlined}[b][.95\columnwidth]<\Bigg(2\big(\log e+\beta_{\alpha,\delta}\log2\big)+\log e\\+2 \log \left( \max_{v \in\supp(Q_V)} \frac{1}{Q_V(v)} \right)\Bigg)n2^{-n\beta_{\alpha,\delta}}\end{multlined}\\
    &\stackrel{(a)}=c_{\alpha,\delta}n2^{-n\beta_{\alpha,\delta}}\numberthis
\end{align*}
where (a) comes from setting
\begin{equation}
c_{\alpha,\delta}\triangleq 3\log e+2\beta_{\alpha,\delta} \log2+2 \log \left( \max_{v \in\supp(Q_V)} \frac{1}{Q_V(v)} \right).\label{EQ:c_alphadelta}
\end{equation}
This implies that for all $\alpha>1$ and $\delta\in\big(0,R-I(U;V)\big)$,
\begin{align*}
     &\mathbb{P}\bigg(D\Big(P^{(\mathbb{B}_n)}_{\mathbf{V}}\Big|\Big|Q_V^n\Big)\geq c_{\alpha,\delta}n2^{-n\beta_{\alpha,\delta}}\bigg)\\
     &\begin{multlined}[b][.95\columnwidth]\leq\mathbb{P}\Bigg(h\left(\int dP_{\mathbb{B}_n,1}\right)+\int dP_{\mathbb{B}_n,1}\log\Delta_{\mathbb{B}_n,1}\\+\int dP_{\mathbb{B}_n,2}\log\Delta_{\mathbb{B}_n,2}\geq c_{\alpha,\delta}n2^{-n\beta_{\alpha,\delta}}\Bigg)\end{multlined}\\
     &\leq \mathbb{P}\big(\mathbb{B}_n\notin\mathcal{S}\big)\\
     &\stackrel{(a)}\leq \left(1+|\mathcal{V}|^n\right)e^{-\frac{1}{3} 2^{n\delta}},\numberthis\label{EQ:bad_divergence_final_UB}
\end{align*}
where (a) follows from \eqref{EQ:notin_S_probability_UB}. Denoting $c_\delta\triangleq \sup_{\alpha> 1}c_{\alpha,\delta}$, \eqref{EQ:bad_divergence_final_UB} further gives
\begin{equation}
\mathbb{P}\bigg(D\Big(P^{(\mathbb{B}_n)}_{\mathbf{V}}\Big|\Big|Q_V^n\Big)\geq c_\delta n2^{-n\beta_{\alpha,\delta}}\bigg)\leq \left(1+|\mathcal{V}|^n\right)e^{-\frac{1}{3} 2^{n\delta}}.\label{EQ:bad_divergence_final_UB_final}
\end{equation}
Since \eqref{EQ:bad_divergence_final_UB_final} is true for all $\alpha>1$, it must also be true, with strict inequality in the LHS, when replacing $\beta_{\alpha,\delta}$ with
\begin{equation}
  \gamma_{\delta}\triangleq\sup_{\alpha> 1}\beta_{\alpha,\delta}= \sup_{\alpha> 1} \frac{\alpha - 1}{2 \alpha - 1} \big(R - \delta - d_{\alpha} (Q_{U,V},Q_U Q_V)\big), \label{EQ:soft_covering_exponent_def}
\end{equation}
which is the exponential rate of convergence stated in \eqref{EQ:soft_covering_exponent} that we derive for the strong soft-covering lemma. This establishes the statement from \eqref{EQ:soft_covering_precise} and proves Lemma \ref{LEMMA:soft_covering}.

Concluding, if $R > I(U;V)$ and for any $\delta\in\big(0,R-I(U;V)\big)$, we get exponential convergence of the relative entropy at rate $O(2^{-\gamma_\delta  n})$ with doubly-exponential certainty. Discarding the precise exponents of convergence and coefficients, we state that there exist $\gamma_1,\gamma_2>0$, such that for $n$ large enough
\begin{equation}
    \mathbb{P}\bigg(D\Big(P^{(\mathbb{B}_n)}_{\mathbf{V}}\Big|\Big|Q_V^n\Big)>e^{-n\gamma_1}\bigg)\leq e^{-e^{n\gamma_2}}.
\end{equation}
\end{IEEEproof}

\section{Wiretap Channel I}\label{SEC:wiretapI}

As a rather simple application of stronger soft-covering lemma, we give an alternative derivation of the SS-capacity of the WTC I \cite{Vardy_Semantic_WTC2012,Ling_SS_Gaussian_WTC2014,Tyagi_Gaussian_WTC2014,Hayashi_SS_BCConfidential2015}. Since the channel to the legitimate user is the same in both WTCs I and II, the maximal error probability analysis presented here is subsequently used to establish reliability for the WTC II.

Our direct proof relies on classic wiretap codes and SS is established using the union bound while invoking the stronger soft-covering lemma. In a wiretap code, a subcode is associated with each confidential message. To transmit a certain message, a codeword from its subcode is selected uniformly at random and transmitted over the channel. Letting these subcodes be large enough while noting that the number of confidential messages only grows exponentially with the blocklength, the union bound and the double-exponential decay the lemma provides show the existence of a semantically-secure sequence of codes. Using these codes, each transmitted message induces an output PMF at the eavesdropper that appears i.i.d. and does not depend on the message.

Wyner's soft-covering lemma, that is now a standard tool for achieving strong-secrecy for the WTC I, comes up short in providing SS. The classic soft-covering argument says that on average over the messages, the output at the eavesdropper will look i.i.d., provided that the size of these subcodes is large enough. This can be used to claim that the unnormalized mutual information between the message and the eavesdropper's output is small, thus ensuring strong-secrecy. However, for SS, it must be claimed that the output PMF is close the i.i.d. distribution for all messages, and there are exponentially many messages. Here is where the stronger soft-covering lemma is advantageous.


\begin{figure}[!t]
    \begin{center}
        \begin{psfrags}
            \psfragscanon
            \psfrag{I}[][][1]{\ \ \ $m$}
            \psfrag{J}[][][1]{\ \ \ \ \ \ \  Encoder}
            \psfrag{X}[][][1]{\ \ \ \ \ \ \ \ Channel}
            \psfrag{K}[][][1]{\ \ \ \ $\mathbf{X}$}
            \psfrag{L}[][][1]{\ \ \ \ \ \ \ \ \ $Q_{Y,Z|X}$}
            \psfrag{S}[][][1]{\ \ \ \ \ \ \ \ \ Channel}
            \psfrag{M}[][][1]{\ \ \ $\mathbf{Y}$}
            \psfrag{N}[][][1]{\ \ \ $\mathbf{Z}$}
            \psfrag{O}[][][1]{\ \ \ \ \ \ Decoder}
            \psfrag{P}[][][0.9]{\ \ \ \ \ \ \ \ Eave.}
            \psfrag{Q}[][][1]{\ \ \ \ \ \ \ $\hat{m}$}
            \psfrag{U}[][][1]{\ $m$}
            \includegraphics[scale = .33]{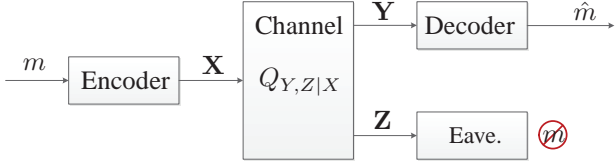}
            \caption{The classic wiretap channel, referred to as the WTC I.} \label{FIG:wiretapI}
            \psfragscanoff
        \end{psfrags}
     \end{center}
 \end{figure}




\subsection{Problem Definition}\label{SUBSEC:WTCI_definition}

The DM-WTC I is illustrated in Fig. \ref{FIG:wiretapI}. The sender chooses a message $m$ from the set $\big[1:2^{nR}\big]$ and maps it into a sequence $\mathbf{x}\in\mathcal{X}^n$ (the mapping may be random). The sequence $\mathbf{x}$ is transmitted over the DM-WTC I with transition probability $Q_{Y,Z|X}$. The output sequences $\mathbf{y}\in\mathcal{Y}^n$ and $\mathbf{z}\in\mathcal{Z}^n$ are observed by the receiver and the eavesdropper, respectively. Based on $\mathbf{y}$, the receiver produces an estimate $\hat{m}$ of $m$. The eavesdropper tries to glean whatever it can about the message from $\mathbf{z}$.

\begin{definition}[Code Description]\label{DEF:WTCI_code}
An $(n,R)$ WTC I code $\mathcal{C}_{n}$ has: 
\begin{enumerate}
\item A message set $\mathcal{M}=\big[1:2^{nR}\big]$.
\item A stochastic encoder $f_1:\mathcal{M}\to\mathcal{P}(\mathbb{X}^n)$.
\item A decoding function $\phi_1: \mathcal{Y}^n\to \hat{\mathcal{M}}$, where $\hat{\mathcal{M}}=\mathcal{M}\cup\{e\}$ and $e\notin\mathcal{M}$.
\end{enumerate}
\end{definition}

For any message distribution $P_M\in\mathcal{P}(\mathcal{M})$, the joint PMF over $\mathcal{M}\times\mathcal{X}^n\times\mathcal{Y}^n\times\mathcal{Z}^n\times\hat{\mathcal{M}}$ induced by $P_M$ and an $(n,R)$ code $\mathcal{C}_{n}$ is:
\begin{align*}
	 P^{(\mathcal{C}_{n})}(m,\mathbf{x},\mathbf{y},\mathbf{z},\hat{m})&=P_M(m)f(\mathbf{x}|m)\\&\mspace{7mu}\times Q_{Y,Z|X}(\mathbf{y},\mathbf{z}|\mathbf{x})\mathds{1}_{\big\{\hat{m}=\phi_1(\mathbf{y})\big\}}.\numberthis\label{EQ:WTCI_induced_PMF}
\end{align*}

\begin{definition}[Maximal Error Probability]\label{DEF:WTCI_error_probability} The maximal error probability of an $(n,R)$ WTC I code $\mathcal{C}_{n}$ is
\begin{subequations}
\begin{equation}
e^\star(\mathcal{C}_{n})=\max_{m\in\mathcal{M}}e_m(\mathcal{C}_{n}),\label{EQ:WTCI_error_prob}
\end{equation}
where
\begin{equation}
	e_m(\mathcal{C}_{n})=\sum_{\mathbf{x}\in\mathcal{X}^n}f_1(\mathbf{x}|m)\sum_{\substack{\mathbf{y}\in\mathcal{Y}^n:\\\phi_1(\mathbf{y})\neq m}}Q_{Y|X}^n(\mathbf{y}|\mathbf{x}).
\end{equation}
\end{subequations}
\end{definition}

\begin{definition}[SS Metric] The SS metric associated to an $(n,R)$ WTC I code $\mathcal{C}_{n}$ is \footnote{$\mathrm{Sem}(\mathcal{C}_{n})$ is actually the mutual-information-security (MIS) metric, which is equivalent to SS by \cite{Vardy_Semantic_WTC2012}. We use the representation in \eqref{EQ:WTCI_security_metric} rather than the formal definition of SS (see, e.g., \cite[Equation (4)]{Vardy_Semantic_WTC2012}) out of analytical convenience.}
\begin{equation}
\mathrm{Sem}(\mathcal{C}_{n})=\max_{P_M\in\mathcal{P}(\mathcal{M})}I_{\mathcal{C}_{n}}(M;\mathbf{Z}),\label{EQ:WTCI_security_metric}
\end{equation}
where $I_{\mathcal{C}_{n}}$ denotes a mutual information term that is calculated with respect to the PMF induced by $\mathcal{C}_{n}$ from \eqref{EQ:WTCI_induced_PMF}.
\end{definition}



\begin{definition}[Semantically-Secure Codes]\label{DEF:SS_codes} A sequence of $(n,R)$ WTC I codes $\big\{\mathcal{C}_{n}\big\}_{n\in\mathbb{N}}$ is semantically-secure if there is a constants $\gamma>0$ and an $n_0\in\mathbb{N}$, such that for every $n>n_0$, $\mathrm{Sem}(\mathcal{C}_{n})\leq
e^{-n\gamma}$.
\end{definition}
%

\begin{remark}\label{REM:Message_PMF_function_of_code} SS requires that a single sequence of codes works well for all message PMFs. Accordingly, the mutual information term in \eqref{EQ:WTCI_security_metric} is maximized over $P_M$ when the code $\mathcal{C}_{n}$ is known. In other words, although not stated explicitly, $P_M$ is a function of $\mathcal{C}_{n}$.
\end{remark}

\begin{remark}
By Definition \ref{DEF:SS_codes}, for a sequence of WTC I codes to be semantically-secure, the SS metric from \eqref{EQ:WTCI_security_metric} must vanish exponentially fast. This is a standard requirement in the cryptography community, commonly referred to as strong-SS (see, e.g., \cite[Section 3.2]{Vardy_Semantic_WTC2012}). The coding scheme given in the direct proof of Theorem \ref{TM:WTCI_capacity_WTC} achieves this exponential decay of the SS-metric (see Section \ref{SUBSUBSEC:WTCI_proof_theorem1}). An exponential decay of the strong-secrecy metric was previously observed in \cite{Hayashi_Secrecy_Resolvability2006,Hayashi_Exponential_Resolvability2011,Hayashi_SS_BCConfidential2015}.
\end{remark}

\begin{definition}[SS-Achievability] A rate $R\in\mathbb{R}_+$ is SS-achievable if there is a sequence of $(n,R)$ WTC I semantically-secure codes $\big\{\mathcal{C}_{n}\big\}_{n\in\mathbb{N}}$ with $e^\star(\mathcal{C}_{n})\to 0$ as $n\to\infty$.
\end{definition}

\begin{definition}[SS-Capacity]
The SS-capacity of the WTC I, $C_{\mathrm{Sem}}$, is the supremum of the set of SS-achievable rates.
\end{definition}

\subsection{Results}\label{SEC:WTCI_result}

\par As stated in the following theorem, the SS-capacity of the WTC I under a maximal error probability constraint is the same as its weak-secrecy-capacity under an average error probability constraint.

\begin{theorem}[WTC I SS-Capacity]\label{TM:WTCI_capacity_WTC}
The SS-capacity of the WTC I is
\begin{equation}
C_{\mathrm{Sem}}=\max_{\substack{Q_{U,X}:\\U-X-(Y,Z)}}\Big[I(U;Y)-I(U;Z)\Big],\label{EQ:WTCI_capacity_WTC}
\end{equation}
and one may restrict the cardinality of $V$ to $|\mathcal{U}|<|\mathcal{X}|$.
\end{theorem}

The proof of Theorem \ref{TM:WTCI_capacity_WTC} is given in Section \ref{SUBSUBSEC:WTCI_proof_theorem1}. Our achievability proof relies on the stronger soft-covering lemma to establish the existence of a sequence of semantically-secure codes with a vanishing average probability of error. The expurgation technique \cite[Theorem 7.7.1]{Cover_Thomas} is then used to upgrade the codes to have a vanishing maximal error probability.

\begin{remark}
The cardinality bound in Theorem \ref{TM:WTCI_capacity_WTC} was established in \cite[Theorem 22.1]{ElGamal2011}.
\end{remark}

\begin{remark}\label{REM:WTCI_SScapacity_no_lemma}
The direct part of Theorem \ref{TM:WTCI_capacity_WTC} can also be derived without using the stronger soft-covering lemma. Instead, one may invoke the codebook expurgation technique twice. By removing a certain portion of the messages, any sequence of codes that ensures strong-secrecy and a vanishing average error probability, can be upgraded to provide SS and reliability with respect to the maximal error probability with negligible rate-loss. In the original codes, the fraction of messages that induce an error probability greater than three times the average, is less than $\frac{1}{3}$. Similarly, the fraction of messages with secrecy distance greater than three times the average is less than $\frac{1}{3}$. Therefore, the fraction of offending messages is less than $\frac{2}{3}$. By removing them one obtains a new sequence of codes that is semantically-secure and has a vanishing maximal error probability. Finally, the rate of the $n$-th code in the new sequence is $R-\frac{\log(3)}{n}$ (here $R$ stands for the rate of the original codes), and the loss is negligible for large $n$.
\end{remark}

\begin{remark}
The expurgation method is insufficient for establishing SS for the WTC II because the messages that need to be removed might differ from one choice of the eavesdropper's observations to the next. It also does not work in other settings such as the multiple access WTC, where expurgation is problematic in general. On the other hand, even for that setting, an achievability proof that relies on the stronger soft-covering lemma goes through by similar steps to those presented below. Thus, strong-secrecy can be upgraded to SS even in situations where vanishing average error probability cannot be upgraded to vanishing maximum error probability (via expurgation).
\end{remark}

\subsection{Proofs}\label{SUBSEC:WTCI_proofs}


\subsubsection{Theorem \ref{TM:WTCI_capacity_WTC}}\label{SUBSUBSEC:WTCI_proof_theorem1}

For the converse, let $\big\{\mathcal{C}_{n}\big\}_{n\in\mathbb{N}}$ be a sequence of $(n,R)$ semantically-secure WTC I codes with $e^\star(\mathcal{C}_{n})\to 0$. Since both $e^\star(\mathcal{C}_{n})\to 0$ and $\mathrm{Sem}(\mathcal{C}_{n})\to 0$ hold for any message distribution $P_M\in\mathcal{P}(\mathcal{M})$, in particular, they hold for a uniform $P_M$. The converse thus follows since $C_{\mathrm{Sem}}$ in \eqref{EQ:WTCI_capacity_WTC} coincides with the secrecy-capacity of the WTC I under a vanishing average error probability criterion and the weak-secrecy constraint.

For the direct part, we first establish the achievability of \eqref{EQ:WTCI_capacity_WTC} when $U=X$. Then, a standard channel prefixing argument extends the proof to any $U$ with $U-X-Y$.

Fix $\epsilon>0$, a PMF $Q_X\in\mathcal{P}(\mathcal{X})$, and let $M$ and $W$ be independent random variables uniformly distributed over $\mathcal{M}$ and $\mathcal{W}\triangleq\big[1:2^{n\tilde{R}}\big]$, respectively. $M$ represents the choice of the message, while $W$ stands for the stochastic part of the encoder. Thus, we start by imposing a uniform distribution over the set of messages and use this to show the existence of a semantically-secure sequence of $(n,R)$ codes with a vanishing \emph{average} error probability. Afterwards, the uniform message distribution assumption is dropped using the expurgation technique \cite[Theorem 7.7.1]{Cover_Thomas}, which allows upgrading reliability to achieve a vanishing \emph{maximal} error probability, while preserving SS.


\par\textbf{Codebook $\bm{\mathcal{B}_n}$:} Let $\mathbb{B}_n$ be a \emph{random codebook} given by a collection of i.i.d. random vectors $\mathbb{B}_n=\big\{\mathbf{X}(m,w)\big\}_{(m,w)\in\mathcal{M}\times\mathcal{W}}$, each distributed according to $Q_X^n$. A realization of $\mathbb{B}_n$ is denoted by $\mathcal{B}_n\triangleq\big\{\mathbf{x}(m,w,\mathcal{B}_n)\big\}_{(m,w)\in\mathcal{M}\times\mathcal{W}}$, with respect to which a classic wiretap code is constructed. 

\par\textbf{Encoder $\bm{f_1}$:} To send $m\in\mathcal{M}$ the encoder randomly and uniformly chooses $W=w$ from $\mathcal{W}$ and transmits $\mathbf{x}(m,w,\mathcal{B}_n)$ over the WTC I. 


\par\textbf{Decoder $\bm{\phi_1}$:} Upon observing $\mathbf{y}\in\mathcal{Y}^n$, the decoder searches for a unique pair $(\hat{m},\hat{w})\in\mathcal{M}\times\mathcal{W}$ such that
\begin{equation}
\big(\mathbf{x}(\hat{m},\hat{w},\mathcal{B}_n),\mathbf{y}\big)\in\mathcal{T}_\epsilon^{n}(Q_{X,Y}).\label{EQ:WTCI_decoding_test}
\end{equation}
If such a unique pair is found, then set $\phi_1(\mathbf{y})=\hat{m}$; otherwise, $\phi_1(\mathbf{y})=e$.

\par The triple $(\mathcal{M},f_1,\phi_1)$ defined with respect to the codebook $\mathcal{B}_n$ constitutes an $(n,R)$ WTC I code $\mathcal{C}_n$. When a random codebook $\mathbb{B}_n$ is used, we denote the corresponding random code by $\mathbb{C}_n$.


\par \textbf{Average Error Probability Analysis:} By standard joint typicality arguments we show that the average error probability, when expected over the ensemble of codebooks, is arbitrarily small. For every fixed codebook  $\mathcal{B}_n$ and $(\tilde{m},\tilde{w})\in\mathcal{M}\times\mathcal{W}$, define the event
\begin{equation}
\mathcal{E}(\tilde{m},\tilde{w},\mathcal{B}_n)=\Big\{\big(\mathbf{x}(\tilde{m},\tilde{w},\mathcal{B}_n),\mathbf{Y}\big)\in\mathcal{T}_\epsilon^{n}(Q_{X,Y})\Big\},
\end{equation}
where $\mathbf{Y}\sim Q^n_{Y|X=\mathbf{x}(\tilde{m},\tilde{w},\mathcal{B}_n)}$ is the random sequence observed at the receiver when the transmitted sends $(\tilde{m},\tilde{w})$. We have
\begin{align*}
&\mathbb{E}_{\mathbb{C}_n}\frac{1}{|\mathcal{M}|}\sum_{m\in\mathcal{M}}e_m(\mathbb{C}_n)\\
&=\mathbb{E}_{\mathbb{C}_n}\mathbb{P}_{\mathbb{C}_n}\big(\hat{M}\neq M\big)\\
&\leq\mathbb{E}_{\mathbb{C}_n}\mathbb{P}_{\mathbb{C}_n}\big((\hat{M},\hat{W})\neq (M,W)\big)\\
&\stackrel{(a)}=\mathbb{E}_{\mathbb{C}_n}\mathbb{P}_{\mathbb{C}_n}\big((\hat{M},\hat{W})\neq (1,1)\big|M=1,W=1\big)\\
&\stackrel{(b)}=\mathbb{E}_{\mathbb{B}_n}\mathbb{P}\left(\mathcal{E}(1,1,\mathbb{B}_n)^c\cup\left\{\bigcup_{(\tilde{m},\tilde{w})\neq(1,1)}\mathcal{E}(\tilde{m},\tilde{w},\mathbb{B}_n)\right\}\vasti|\mspace{2mu}\mathbb{B}_n\right)\\
&\begin{multlined}[b][.9\columnwidth]\stackrel{(c)}\leq \underbrace{\mathbb{P}_{Q^n_{X,Y}}\Big((\mathbf{X},\mathbf{Y})\in\mathcal{T}_\epsilon^{n}(Q_{X,Y})\Big)}_{P_1}\\+\underbrace{\sum_{(\tilde{m},\tilde{w})\neq(1,1)}\mathbb{P}_{Q^n_X\times Q^n_Y}\Big((\mathbf{X},\mathbf{Y})\in\mathcal{T}_\epsilon^{n}(Q_{X,Y})\Big)}_{P_2}\end{multlined},\numberthis
\end{align*}
where (a) uses the symmetry of the codebook construction with respect to $m$ and $w$, (b) follows by the decoding rule, while (c) takes the expectation over the ensemble of codebooks and uses the union bound.

By the law of large numbers $P_1\to 0$ as $n\to\infty$, while $P_2\to 0$ as $n$ grows provided that\footnote{All subsequent mutual information terms in the proof are calculated with respect to $Q_{U,X}Q_{Y,Z|X}$ or its marginals.}
\begin{equation}
R+\tilde{R}<I(X;Y).\label{EQ:WTCI_reliability_constraint}
\end{equation}
Thus, we have
\begin{equation}
\mathbb{E}_{\mathbb{C}_n}\frac{1}{|\mathcal{M}|}\sum_{m\in\mathcal{M}}e_m(\mathbb{C}_n)\xrightarrow[n\to\infty]{}0.\label{EQ:WTCI_average_error_prob_vanish}
\end{equation}


\par \textbf{Security Analysis:} 
For any fixed $\mathcal{B}_n$ (which, in turn, fixed $\mathcal{C}_n$), we denote by $P^{(\mathcal{C}_n)}_{M,\mathbf{Z}}$ the joint distribution of $M$ and $\mathbf{Z}$ induced by the code $\mathcal{C}_n$ (see \eqref{EQ:WTCI_induced_PMF}). For any $\mathcal{B}_n$, we first have
\begin{align*}
\max_{P_M\in\mathcal{P}(\mathcal{M})}&I_{\mathcal{C}_n}(M;\mathbf{Z})\\
&\stackrel{(a)}=\max_{P_M\in\mathcal{P}(\mathcal{M})}D\Big(P^{(\mathcal{C}_n)}_{\mathbf{Z}|M}\Big|\Big|P^{(\mathcal{C}_n)}_{\mathbf{Z}}\Big|P_{M}\Big)\\
&\stackrel{(b)}\leq\max_{P_M\in\mathcal{P}(\mathcal{M})}D\Big(P^{(\mathcal{C}_n)}_{\mathbf{Z}|M}\Big|\Big|Q_Z^n\Big|P_{M}\Big)\\
&=\max_{P_M\in\mathcal{P}(\mathcal{M})}\sum_{m\in\mathcal{M}}P(m)D\Big(P^{(\mathcal{C}_n)}_{\mathbf{Z}|M=m}\Big|\Big|Q_Z^n\Big)\\
&\leq\max_{P_M\in\mathcal{P}(\mathcal{M})}\sum_{m\in\mathcal{M}}P(m)\max_{\tilde{m}\in\mathcal{M}}D\Big(P^{(\mathcal{C}_n)}_{\mathbf{Z}|M=\tilde{m}}\Big|\Big|Q_Z^n\Big)\\
&=\max_{m\in\mathcal{M}}D\Big(P^{(\mathcal{C}_n)}_{\mathbf{Z}|M=m}\Big|\Big|Q_Z^n\Big),\numberthis\label{EQ:WTCI_proof_prop1}
\end{align*}
where (a) uses the relative entropy chain rule, while is because for any $P_M\in\mathcal{P}(\mathcal{M})$, we have
\begin{align*}
&D\Big(P^{(\mathcal{C}_n)}_{\mathbf{Z}|M}\Big|\Big|P^{(\mathcal{C}_n)}_{\mathbf{Z}}\Big|P_{M}\Big)\\
&=\sum_{m\in\mathcal{M}}P(m)\sum_{\mathbf{z}\in\mathcal{Z}^n}P^{(\mathcal{C}_n)}(\mathbf{z}|m)\log\left(\frac{P^{(\mathcal{C}_n)}(\mathbf{z}|m)}{P^{(\mathcal{C}_n)}(\mathbf{z}\big)}\cdot\frac{Q_Z^n(\mathbf{z})}{Q_Z^n(\mathbf{z})}\right)\\
      &\begin{multlined}[b][.95\columnwidth]=D\Big(P^{(\mathcal{C}_n)}_{\mathbf{Z}|M}\Big|\Big|Q_Z^n\Big|P_M\Big)\\-\sum_{m\in\mathcal{M}}P(m)\sum_{\mathbf{z}\in\mathcal{Z}^n}P^{(\mathcal{C}_n)}(\mathbf{z}|m)\log\left(\frac{P^{(\mathcal{C}_n)}(\mathbf{z})}{Q_Z^n(\mathbf{z})}\right)\end{multlined}\\
      &=D\Big(P^{(\mathcal{C}_n)}_{\mathbf{Z}|M}\Big|\Big|Q_Z^n\Big|P_M\Big)-D\Big(P^{(\mathcal{C}_n)}_{\mathbf{Z}}\Big|\Big|Q_Z^n\Big)\\
      &\leq D\Big(P^{(\mathcal{C}_n)}_{\mathbf{Z}|M}\Big|\Big|Q_Z^n\Big|P_M\Big)\numberthis\label{EQ:WTCI_divergence_grows}.
\end{align*}

Now, let $\tilde{\gamma}$ be an arbitrary positive real number to be determined later and consider the following probability.
\begin{align*}
\mathbb{P}\bigg(&\Big\{\mathrm{Sem}(\mathbb{C}_n)\leq e^{-n\tilde{\gamma}}\Big\}^c\bigg)\\&\stackrel{(a)}\leq\mathbb{P}\bigg(\Big\{\max_{m\in\mathcal{M}} D\Big(P^{(\mathbb{C}_n)}_{\mathbf{Z}|M=m}\Big|\Big|Q_Z^n\Big)\leq e^{-n\tilde{\gamma}}\Big\}^c\bigg)\\
&=\mathbb{P}\bigg(\Big\{\forall m\in\mathcal{M},\ D\Big(P^{(\mathbb{C}_n)}_{\mathbf{Z}|M=m}\Big|\Big|Q_Z^n\Big)\leq e^{-n\tilde{\gamma}}\Big\}^c\bigg)\\
&=\mathbb{P}\bigg(\exists m\in\mathcal{M},\ D\Big(P^{(\mathbb{C}_n)}_{\mathbf{Z}|M=m}\Big|\Big|Q_Z^n\Big)> e^{-n\tilde{\gamma}}\bigg)\\
&=\mathbb{P}\Bigg(\bigcup_{m\in\mathcal{M}}\Big\{D\Big(P^{(\mathbb{C}_n)}_{\mathbf{Z}|M=m}\Big|\Big|Q_Z^n\Big)> e^{-n\tilde{\gamma}}\Big\}\Bigg)\\
&\leq\sum_{m\in\mathcal{M}}\mathbb{P}\bigg(D\Big(P^{(\mathbb{C}_n)}_{\mathbf{Z}|M=m}\Big|\Big|Q_Z^n\Big)> e^{-n\tilde{\gamma}}\bigg),\numberthis\label{EQ:WTCI_secrecy_comp_probability_UB}
\end{align*}
where (a) follows from \eqref{EQ:WTCI_proof_prop1} and \eqref{EQ:WTCI_proof_prop1}.

By the stronger soft-covering lemma, if
\begin{equation}
\tilde{R}>I(X;Z)\label{EQ:WTCI_secrecy_constraint},
\end{equation}
then there are $\gamma_1,\gamma_2>$ such that
\begin{equation}
\mathbb{P}\bigg(D\Big(P^{(\mathbb{C}_n)}_{\mathbf{Z}|M=m}\Big|\Big|Q_Z^n\Big)> e^{-n\gamma_1}\bigg)\leq e^{-e^{n\gamma_2}},\label{EQ:WTCI_secrecy_soft_covering_result}
\end{equation}
for sufficiently large $n$. Inserting (\ref{EQ:WTCI_secrecy_soft_covering_result}) into (\ref{EQ:WTCI_secrecy_comp_probability_UB}) while setting $\tilde{\gamma}=\gamma_1$, we have
\begin{align*}
\mathbb{P}\bigg(\Big\{\mathrm{Sem}(\mathbb{C}_n)\leq e^{-n\gamma_1}\Big\}^c\bigg)&\leq \sum_{m\in\mathcal{M}}e^{-e^{n\gamma_2}}\\
&=2^{nR}\cdot e^{-e^{n\gamma_2}}\\
&\triangleq \eta_n\xrightarrow[n\to\infty]{}0\numberthis\label{EQ:WTCI_secrecy_comp_probability_final},
\end{align*}
and therefore,
\begin{equation}
\mathbb{P}\Big(\mathrm{Sem}(\mathbb{C}_n)\leq e^{-n\gamma_1}\Big)\geq 1-\eta_n\xrightarrow[n\to\infty]{}1\label{EQ:WTCI_secrecy_probability_final}.
\end{equation}
Inequality (\ref{EQ:WTCI_secrecy_probability_final}) implies that if $\tilde{R}$ satisfies (\ref{EQ:WTCI_secrecy_constraint}), the probability that a randomly generated sequence of codes meets the SS criterion for large $n$ is arbitrarily close to 1. In fact, because \eqref{EQ:WTCI_secrecy_comp_probability_final} decays so rapidly, the Borel-Cantelli lemma implies that almost every sequence of realizations of $\big\{\mathbb{C}_n\big\}_{n\in\mathbb{N}}$ is semantically-secure.


\par \textbf{SS-Achievability:} To establish the existence of a sequence of $(n,2^{nR})$ reliable and semantically-secure codes $\big\{\mathcal{C}_n\big\}_{n\in\mathbb{N}}$, we reproduce the Selection Lemma \cite[Lemma 2.2]{Bloch_Barros_Secrecy_Book2011}.
\begin{lemma}[Selection Lemma]\label{LEMMA:WTCI_selection}
Let $\big\{A_n\big\}_{n\in\mathbb{N}}$ be a sequence of random variables, where $A_n$ takes values in $\mathcal{A}_n$. Let $\big\{f_n^{(1)},f_n^{(2)},\ldots,f_n^{(I)}\big\}_{n\in\mathbb{N}}$ be a collection of $I<\infty$ sequences of bounded functions $f_n^{(i)}:\mathcal{A}_n\to\mathbb{R}_+$, $i\in[1:I]$. If
\begin{subequations}
\begin{equation}
\mathbb{E}f_n^{(i)}(A_n)\xrightarrow[n\to\infty]{}0,\quad\forall i\in[1:I],\label{EQ:WTCI_selection_it}
\end{equation}
then there exists a sequence $\{a_n\}_{n\in\mathbb{N}}$, where $a_n\in\mathcal{A}_n$ for every $n\in\mathbb{N}$, such that
\begin{equation}
f_n^{(i)}(a_n)\xrightarrow[n\to\infty]{}0,\quad\forall i\in[1:I].\label{EQ:WTCI_selection_then}
\end{equation}
\end{subequations}
\end{lemma}

For completeness, the proof of Lemma \ref{LEMMA:WTCI_selection} is given in Appendix \ref{APPEN:selection_proof}. Applying Lemma \ref{LEMMA:WTCI_selection} to the random variables $\big\{\mathbb{C}_n\big\}_{n\in\mathbb{N}}$ and the functions $\frac{1}{|\mathcal{M}|}\sum_{m\in\mathcal{M}}e_m(\mathbb{C}_n)$ and $\mathds{1}_{\big\{\mathrm{Sem}(\mathbb{C}_n)>e^{-n\gamma_1}\big\}}$, while using \eqref{EQ:WTCI_average_error_prob_vanish} and \eqref{EQ:WTCI_secrecy_comp_probability_final}, we have that there is a sequence of $(n,R)$ WTC I codes $\big\{\mathcal{C}_n\big\}_{n\in\mathbb{N}}$, for which
\begin{subequations}
\begin{align}
&\frac{1}{|\mathcal{M}|}\sum_{m\in\mathcal{M}}e_m(\mathcal{C}_n)\xrightarrow[n\to\infty]{}0,\label{EQ:WTCI_average_error_prob_vanish_fixed_code}\\
&\mathds{1}_{\big\{\mathrm{Sem}(\mathcal{C}_n)>e^{-n\gamma_1}\big\}}\xrightarrow[n\to\infty]{}0\label{EQ:WTCI_indicator_secrecy_vanish_fixed_code}.
\end{align}
\end{subequations}
Since the indicator function in (\ref{EQ:WTCI_indicator_secrecy_vanish_fixed_code}) takes only the values 0 and 1, to satisfy the convergence there must exist an $n_0\in\mathbb{N}$, such that
\begin{equation}
\mathds{1}_{\big\{\mathrm{Sem}(\mathcal{C}_n)>e^{-n\gamma_1}\big\}}=0,\quad\forall n>n_0,\label{EQ:WTCI_indicator_secrecy_zero}
\end{equation}
and therefore,
\begin{equation}
\mathrm{Sem}(\mathcal{C}_n)\leq e^{-n\gamma_1},\quad\forall n>n_0.\label{EQ:WTCI_S_secrecy_achieved}
\end{equation}

The final step is to amend $\big\{\mathcal{C}_n\big\}_{n\in\mathbb{N}}$ to be reliable with respect to the maximal error probability (as defined in \eqref{EQ:WTCI_error_prob}). This is done using the expurgation technique (see, e.g., \cite[Theorem 7.7.1]{Cover_Thomas}). Namely, we discard the worst half of the codewords in each codebook $\mathcal{B}_n$. Denoting the amended sequence of codebooks by $\big\{\mathcal{B}^\star_n\big\}_{n\in\mathbb{N}}$ and their corresponding sequence of codes by $\big\{\mathcal{C}^\star_n\big\}_{n\in\mathbb{N}}$, we have
\begin{equation}
e^\star(\mathcal{C}^\star_n)\xrightarrow[n\to\infty]{}0.
\end{equation}
Note that in each $\mathcal{C}^\star_n$ there are $2^{nR-1}$ codewords, i.e., throwing out half the codewords has changed the rate from $R$ to $R-\frac{1}{n}$, which is negligible for large $n$. Further note that because $\big\{\mathcal{C}_n\big\}_{n\in\mathbb{N}}$ is semantically-secure, so is $\big\{\mathcal{C}^\star_n\big\}_{n\in\mathbb{N}}$. Combining \eqref{EQ:WTCI_reliability_constraint} with \eqref{EQ:WTCI_secrecy_constraint}, we have that every
\begin{equation}
0\leq R<\max_{Q_X}\Big[I(X;Y)-I(X;Z)\Big]\label{EQ:WTCI_achievable_rate_V=X}
\end{equation}
is SS-achievable.

\par \par To establish the achievability of $C_{\mathrm{Sem}}$ from \eqref{EQ:WTCI_capacity_WTC}, we prefix a DM-channel (DMC) $Q_{X|V}$ to the original WTC I $Q_{Y,Z|X}$ to obtain a new channel $Q_{Y,Z|V}$, where
\begin{equation}
Q^n_{Y,Z|V}(\mathbf{y},\mathbf{z}|\mathbf{v})=\sum_{\mathbf{x}\in\mathcal{X}^n}Q^n_{X|V} (\mathbf{x}|\mathbf{v})Q^n_{Y,Z|X}(\mathbf{y},\mathbf{z}|\mathbf{x}).
\end{equation}
Using a similar analysis as above with respect to $Q_{Y,Z|V}$, any $R\in\mathbb{R}^+$ satisfying
\begin{equation}
R<\max_{\substack{Q_{U,X}:\\U-X-(Y,Z)}}\Big[I(U;Y)-I(U;Z)\Big]\label{EQ:WTCI_achievable_rate_V}
\end{equation}
is achievable.

\section{Wiretap Channel II}\label{SEC:wiretapII}

The WTC II scenario considers communication between two legitimate parties in the presence of an eavesdropper that can choose to observed any subset of the transmitted sequence, while being limited in quantity. The challenge in this setting is that the eavesdropper knows the codebook when it selects the subset to observe. Therefore, secrecy will only be achieved if it is achieved uniformly for all selections of packets, of which there are exponentially many possibilities. Furthermore, SS being our goal, secrecy must be ensured for each one of the exponentially many confidential messages. Nonetheless, as the combined number of subsets and messages grows only exponentially with the blocklenght, using the stronger soft-covering lemma we show that rates all the way up to the weak-secrecy-capacity of the DM erasure WTC I are achievable even in this more stringent setting. Then, we establish the capacity of this WTC I as an upper bound on the considered WTC II, thus characterizing its SS-capacity.


\subsection{Problem Definition}\label{SUBSEC:WTCII_definition}

\begin{figure}[t!]
    \begin{center}
        \begin{psfrags}
            \psfragscanon
            \psfrag{I}[][][1]{\ \ \ \ $m$}
            \psfrag{J}[][][1]{\ \ \ Trans.}
            \psfrag{A}[][][0.9]{\ \ \ \ \ \ \ \ \ \ \ \ \ \ \ \  $\mspace{2mu}\mathcal{S}\mspace{-2mu}\subseteq\mspace{-2mu}[1\mspace{-3mu}:\mspace{-3mu}n], |\mathcal{S}|\mspace{-2mu}=\mspace{-2mu}\mu$}
            \psfrag{B}[][][0.9]{\ \ \ \ \ \ \ \ \ \ \ \ \ \ \ \ \ \ $Z_i\mspace{-2mu}=\mspace{-2mu}\begin{cases}X_i\mspace{2mu},\mspace{3mu}i\in\mathcal{S} \\ ?,\mspace{17mu}i\notin\mathcal{S}\end{cases}$}
            \psfrag{K}[][][1]{\ \ $\mathbf{X}$}
            \psfrag{L}[][][1]{\ \ $Q_{Y|X}$}
            \psfrag{M}[][][1]{\ $\mathbf{Y}$}
            \psfrag{N}[][][1]{$\mathbf{Z}$}
            \psfrag{O}[][][1]{\ \ Rec.}
            \psfrag{P}[][][1]{\ \ \ Eave.}
            \psfrag{Q}[][][1]{\ $\hat{m}$}
            \psfrag{U}[][][1]{\ $m$}
            \includegraphics[scale = .38]{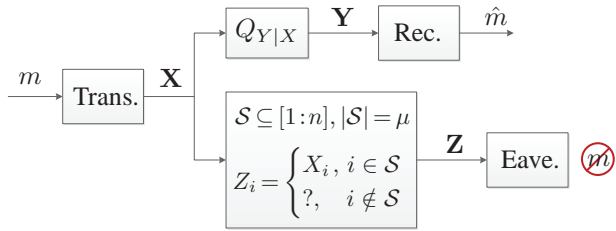}
            \caption{The type II wiretap channel.} \label{FIG:wiretapII}
            \psfragscanoff
        \end{psfrags}
     \end{center}
 \end{figure}

The WTC II is illustrated in Fig. \ref{FIG:wiretapII}. The sender chooses a message $m$ from the set $\big[1:2^{nR}\big]$ and maps it into a sequence $\mathbf{x}\in\mathcal{X}^n$ (the mapping may be random). The sequence $\mathbf{x}$ is transmitted over a point-to-point DMC with transition probability $Q_{Y|X}$. Based on the received channel output sequence $\mathbf{y}\in\mathcal{Y}^n$, the receiver produces an estimate $\hat{m}$ of $m$. The eavesdropper noiselessly observes a subset of its choice of the $n$ transmitted symbols. Namely, the eavesdropped chooses $\mathcal{S}\subseteq[1:n]$, $|\mathcal{S}|=\mu\leq n$, and observes $\mathbf{z}\in\big(\mathcal{X}\cup\{?\}\big)^n$, where
\begin{equation}
z_i=\begin{cases}x_i\mspace{2mu},\ i\in\mathcal{S} \\ ?,\mspace{17mu}i\notin\mathcal{S}\end{cases}.
\end{equation}
Based on $\mathbf{z}$, the eavesdropper tries to learn as much as possible about the message.


With some abuse of notation (reusing notations from Section \ref{SUBSEC:WTCI_definition}), we introduce the following definitions. An $(n,R)$ WTC II code $\mathcal{C}_{n}$ and the corresponding maximal error probability $e^\star(\mathcal{C}_{n})$ are defined similarly to Definitions \ref{DEF:WTCI_code} and \ref{DEF:WTCI_error_probability}, respectively.

\begin{definition}[SS Metric]\label{DEF:WTCII_SS_metric} The SS metric with respect to an $(n,R)$ WTC II code $\mathcal{C}_{n}$ is
\begin{equation}
\mathrm{Sem}_{\mu}(\mathcal{C}_{n})=\max_{\substack{P_M\in\mathcal{P}(\mathcal{M}),\\\mspace{2mu}\mathcal{S}\subseteq[1:n]:\ |\mathcal{S}|=\mu}}I_{\mathcal{C}_{n}}(M;\mathbf{Z}),\label{EQ:WTCII_security_metric}
\end{equation}
where $I_{\mathcal{C}_{n}}$ denotes that the mutual information term is calculated with respect to
\begin{equation*}
P^{(\mathcal{C}_{n},\mathcal{S})}_{M,\mathbf{Z}}(m,\mathbf{z})\mspace{-2mu}=\mspace{-2mu}P(m)\mspace{-8mu}\sum_{\mathbf{x}\in\mathcal{X}^n}\mspace{-5mu}f(\mathbf{x}|m)\mathds{1}_{{\big\{z_i=x_i,\mspace{2mu} i\in\mathcal{S}\big\}}\cap{\big\{z_i=?,\mspace{2mu} i\notin\mathcal{S}\big\}}}\mspace{-2mu}.
\end{equation*}
\end{definition}

\begin{remark} As explained in Remark \ref{REM:Message_PMF_function_of_code}, the code $\mathcal{C}_n$ is known when the mutual information term in \eqref{EQ:WTCII_security_metric} is maximized. Thus, the observed subset $\mathcal{S}\subseteq[1:n]$ and the message PMF $P_M$ are both functions of $\mathcal{C}_n$. Although, for the sake of simplicity, this dependence is omitted from our notations, the reader should keep in mind that a single codebook is required to works well for all choices of subsets and message PMFs.
\end{remark}



\begin{definition}[Semantically-Secure Codes]\label{DEF:SS_metric} Let $\alpha\in[0,1]$ and $\mu=\lfloor\alpha n\rfloor$, a sequence of $(n,R)$ WTC II codes $\big\{\mathcal{C}_n\big\}_{n\in\mathbb{N}}$ is $\alpha$-semantically-secure if there is a constants $\gamma>0$ and an $n_0\in\mathbb{N}$, such that for every $n>n_0$, $\mathrm{Sem}_{\mu}(\mathcal{C}_n)\leq
e^{-n\gamma}$.
\end{definition}


\begin{definition}[SS-Achievability]\label{DEF:achievability_WTCII} Let $\alpha\in[0,1]$ and $\mu=\lfloor\alpha n\rfloor$, a rate $R\in\mathbb{R}_+$ is $\alpha$-SS-achievable if there is a sequence of $(n,R)$ $\alpha$-semantically-secure WTC II codes $\big\{\mathcal{C}_n\big\}_{n\in\mathbb{N}}$ with $e^\star(\mathcal{C}_n)\to 0$ as $n\to\infty$.
\end{definition}

\begin{definition}[SS-Capacity]
For any $\alpha\in[0,1]$, the $\alpha$-SS-capacity of the WTC II $C_{\mathrm{Sem}}(\alpha)$ is the supremum of the set of $\alpha$-SS-achievable rates.
\end{definition}


\subsection{Converse}

The following proposition is subsequently used for the converse proof of the WTC II SS-capacity. The proposition states that the strong-secrecy-capacity of a WTC I with a DM-EC to the eavesdropper is an upper bound on the strong-secrecy-capacity of the WTC II. To formulate the result, slight modifications of some of the definitions from Sections \ref{SUBSEC:WTCI_definition} and \ref{SUBSEC:WTCII_definition} are required. Specifically, we redefine the achievable rates for each setting with respect to a strong-secrecy requirement (instead of SS).

\begin{definition}[Strong-Secrecy Achievability for WTC I]
A rate $R\in\mathbb{R}_+$ is strong-secrecy-achievable for the WTC I if there is a sequence of $(n,R)$ codes $\big\{\mathcal{C}_{1,n}\big\}_{n\in\mathbb{N}}$ with
\begin{subequations}
\begin{align}
e^\star\big(\mathcal{C}_{1,n}\big)&\xrightarrow[n\to\infty]{}0\label{EQ:WTCI_strong_secrecy_error}\\
I_{\mathcal{C}_{1,n}}\big(M;\mathbf{Z}\big)&\xrightarrow[n\to\infty]{}0,\label{EQ:WTCI_strong_secrecy_metric}
\end{align}\label{EQ:WTCI_strong_secrecy_achievability}
\end{subequations}
where $M$ is uniformly distributed over the message set $\mathcal{M}$.

\end{definition}

\begin{definition}[Strong-Secrecy Achievability for WTC II]\label{DEF:WTCII_strong_secrecy_achievability}
Let $\alpha\in[0,1]$ and $\mu=\lfloor\alpha n\rfloor$, a rate $R\in\mathbb{R}_+$ is $\alpha$-strong-secrecy-achievable for the WTC II if there is a sequence of $(n,R)$ codes $\big\{\mathcal{C}_{2,n}\big\}_{n\in\mathbb{N}}$ with
\begin{subequations}
\begin{align}
e^\star\big(\mathcal{C}_{2,n}\big)&\xrightarrow[n\to\infty]{}0\label{EQ:WTCII_strong_secrecy_error}\\
\max_{\substack{\mathcal{S}\subseteq[1:n]:\\|\mathcal{S}|=\mu}}I_{\mathcal{C}_{2,n}}\big(M;\mathbf{Z}\big)&\xrightarrow[n\to\infty]{}0,\label{EQ:WTCII_strong_secrecy_metric}
\end{align}\label{EQ:WTCII_strong_secrecy_achievability}
\end{subequations}
where $M$ is uniformly distributed over the message set $\mathcal{M}$.
\end{definition}

The \emph{strong-secrecy-capacity} for both setting is defined as the supremum of the set of strong-secrecy-achievable rates.

\begin{proposition}[WTC I Upper Bounds WTC II]\label{PROP:WTCI_UB_WTCII}
Let $\alpha\in(0,1]$ and $C_\mathrm{S}^\mathrm{II}(\alpha)$ be the $\alpha$-strong-secrecy-capacity of the WTC II with a main channel $Q^{(2)}_{Y|X}$. Furthermore, let $\beta\in[0,\alpha)$ and $C_\mathrm{S}^\mathrm{I}(\beta)$ be the strong-secrecy-capacity of the WTC I with transition probability $Q^{(1)}_{Y,Z|X}=Q^{(2)}_{Y|X}\mathcal{E}^{(\beta)}_{Z|X}$, where $\mathcal{E}^{(\beta)}_{Z|X}$ is a DM-EC with erasure probability $\bar{\beta}=1-\beta$, i.e.,
\begin{equation}
\mathcal{E}^{(\beta)}_{Z|X}(z|x)=\begin{cases}\beta,\ z=x\\ \bar{\beta},\ z=?\end{cases},\ \forall x\in\mathcal{X}.
\end{equation}
Then
\begin{equation}
C_\mathrm{S}^\mathrm{II}(\alpha)\leq C_\mathrm{S}^\mathrm{I}(\beta)=\max_{\substack{Q_{U,X}:\\U-X-Y}}\Big[I(U;Y)-\beta I(U;X)\Big].\label{EQ:WTCI_UB_WTCII}
\end{equation}
\end{proposition}

See Section \ref{SUBSUBSEC:WTCII_proof_proposition} for the proof. Proposition \ref{PROP:WTCI_UB_WTCII} is subsequently combined with the following lemma to to establish the converse for the $\alpha$-SS-capacity of the WTC II.

\begin{lemma}[Continuity of WTC I Capacity]\label{LEMMA:continues}
As a function of $\beta$,
\begin{equation}
C_\mathrm{S}^\mathrm{I}(\beta)=\max_{\substack{Q_{U,X}:\\U-X-Y}}\Big[I(U;Y)-\beta I(U;X)\Big]
\end{equation}
is continues inside $(0,1)$.
\end{lemma}

The proof of Lemma \ref{LEMMA:continues} is relegated to Appendix \ref{APPEN:contineous_proof}. The SS-capacity of the WTC II with a noisy main channel is stated next.

\begin{theorem}[WTC II SS-Capacity]\label{TM:WTCII_capacity_WTC}
For any $\alpha\in[0,1]$,
\begin{equation}
C_{\mathrm{Sem}}(\alpha)=\max_{\substack{Q_{U,X}:\\U-X-Y}}\Big[I(U;Y)-\alpha I(U;X)\Big],\label{EQ:WTCII_capacity_WTC}
\end{equation}
and one may restrict the cardinality of $U$ to $|\mathcal{U}|<|\mathcal{X}|$.
\end{theorem}

The converse and direct parts of Theorem \ref{TM:WTCII_capacity_WTC} are established in Sections \ref{SUBSUBSEC:WTCII_proof_theorem1_converse} and \ref{SUBSUBSEC:WTCII_proof_theorem1_direct}, respectively. As oppose to the SS-capacity of the WTC I (where achievability may be derived without using Lemma \ref{LEMMA:soft_covering} - see Remark \ref{REM:WTCI_SScapacity_no_lemma}), for the WTC II, the stronger soft-covering lemma is essential for the direct proof. Specifically, via the union bound, the double-exponential decay that Lemma \ref{LEMMA:soft_covering} provides is leveraged to show the existence of a sequence of codes that satisfies the vanishing information leakage requirement for all choices of $\mathcal{S}$ and $P_M$.

\begin{remark}[Generalized WTC II SS-Capacity]
The proof of Theorem \ref{TM:WTCII_capacity_WTC} is robust and readily extends to a more general setting where the  eavesdropper's observed symbols are corrupted by random noise. Specifically, we refer to the scenario where the eavesdropper first chooses a subset of indices $\mathcal{S}\subseteq[1:n]$ of size $\mu=\lfloor \alpha n\rfloor$, then $\mathbf{x}^{\mathcal{S}}$ is passed through a DMC $Q_{Z|X}$ and the eavesdropper receives
$Z_i\sim Q_{Z|X=x_i}$, for $i\in\mathcal{S}$, and $Z=?$ otherwise. The $\alpha$-SS-capacity for this case is
\begin{equation}
C^{(\mathrm{Noisy})}_{\mathrm{Sem}}(\alpha)=\max_{\substack{Q_{U,X}:\\U-X-(Y,Z)}}\Big[I(U;Y)-\alpha I(U;Z)\Big],\label{EQ:WTCIII_capacity_WTC}
\end{equation}
and recovers \eqref{EQ:WTCII_capacity_WTC} by setting $Z=X$. Both the direct and the converse proofs of \eqref{EQ:WTCIII_capacity_WTC} follow by a verbatim repetition of the arguments from Section \ref{SUBSEC:WTCII_proofs}, with two minor changes. First for the converse, the classic DM-EC from Proposition \ref{PROP:WTCI_UB_WTCII} (proven in \ref{SUBSUBSEC:WTCII_proof_proposition}) is replaced with a cascade of the DM-EC and the DMC $Q_{Z|X}$. Second, for the SS analysis in the direct proof (Section \ref{SUBSUBSEC:WTCII_proof_theorem1_direct}) we replace the rate bound from \eqref{EQ:WTCII_prefix_RB2} with $\tilde{R}>\alpha I(U;Z)$ (the reliability analysis goes through without changes). 
\end{remark}

\begin{remark}
The cardinality bound in Theorem \ref{TM:WTCII_capacity_WTC} is established using the convex cover method \cite[Appendix C]{ElGamal2011}. The details are omitted.
\end{remark}

\begin{remark}
Theorem \ref{TM:WTCII_capacity_WTC} recovers the achievability result from \cite[Equation (7)]{Yener_WTCII2015} by setting $U=X$ and taking $X$ to be uniformly distributed over $\mathcal{X}$. Furthermore, in \cite{Yener_WTCII2015} secrecy was established while assuming a uniform distribution over the message set, i.e., on average over the messages. Although we require security with respect to a stricter metric (SS versus weak-secrecy), we achieve higher rates than \cite[Equation (7)]{Yener_WTCII2015} and show their optimality. Moreover, to achieve \eqref{EQ:WTCII_capacity_WTC}, we use classic wiretap codes and establish SS using the stronger soft-covering lemma, making the (rather convoluted) coset coding scheme from \cite{Yener_WTCII2015} (inspired by \cite{Wyner_WTCII1984}) no longer required.
\end{remark}



\subsection{Proofs}\label{SUBSEC:WTCII_proofs}


\subsubsection{Proposition \ref{PROP:WTCI_UB_WTCII}}\label{SUBSUBSEC:WTCII_proof_proposition}

The equality in \eqref{EQ:WTCI_UB_WTCII} follows by evaluating the strong-secrecy-capacity formula of a general WTC I, i.e.,
\begin{equation}
\max_{\substack{Q_{U,X}:\\U-X-(Y,Z)}}\Big[I(U;Y)-I(U;Z)\Big],\label{EQ:WTCI_general_capacity}
\end{equation}
for the case where the transition probability matrix is $Q^{(1)}_{Y,Z|X}=Q^{(2)}_{Y|X}\mathcal{E}^{(\beta)}_{Z|X}$. Let $\Phi\sim\ber(\beta)$ be a random variable, such that its i.i.d. samples define the erasure process of the DM-EC with erasure probability $\bar{\beta}$. Accordingly, $\Phi$ is independent of $X$ and
\begin{equation}
Z=\begin{cases}X,\ \Phi=0\\?,\ \ \mspace{2mu}\Phi=1\end{cases}.
\end{equation}

First note that $\Phi$ is determined by $Z$ since $?\notin\mathcal{X}$. Combining this with the Markov relation $U-X-(Y,Z)$ implies that the chain $U-X-(Y,Z,\Phi)$ is also Markov. Along with the independence of $X$ and $\Phi$, this implies that $U$ and $\Phi$ are also independent. Consequently, for every $Q_{U,X}$, where $U-X-(Y,Z)$ forms a Markov chain, we have
\begin{align*}
I(U;Z)&\stackrel{(a)}=I(U;\Phi,Z)\\
&\stackrel{(b)}=I(U;Z|\Phi)\\
&\stackrel{(c)}=\beta I(U;X)+\bar{\beta}I(U;?)\\
&=\beta I(U;X),\numberthis\label{EQ:V_information_equality}
\end{align*}
where (a) follows since $\Phi$ is defined by $Z$, while (b) and (c) follows by the independence of $\Phi$ and $U$. Since \eqref{EQ:V_information_equality} holds for every $Q_{U,X}$ as above, we conclude that
\begin{equation}
C_\mathrm{S}^\mathrm{I}(\beta)=\max_{\substack{Q_{U,X}:\\U-X-Y}}\Big[I(U;Y)-\beta I(U;X)\Big].\label{EQ:WTCI_general_capacity_evaluated}
\end{equation}

%



To prove the inequality in \eqref{EQ:WTCI_UB_WTCII}, we show that for any $\alpha\in(0,1]$ and $\beta\in[0,\alpha)$, an $\alpha$-strong-secrecy-achievable rate for the WTC II is also achievable for the WTC I with erasure probability $\bar{\beta}$.

Fix $\alpha,\beta$ as above and let $R\in\mathbb{R}_+$ be an $\alpha$-strong-secrecy-achievable rate for the WTC II. Furthermore, let $\big\{\mathcal{C}_{2,n}\big\}_{n\in\mathbb{N}}$ be the corresponding sequence of $(n,R)$ codes satisfying \eqref{EQ:WTCII_strong_secrecy_achievability}. Since the channel to the legitimate receiver and the definition of the maximal error probability are the same for both versions of the WTC (see (\ref{EQ:WTCI_strong_secrecy_error}) and (\ref{EQ:WTCII_strong_secrecy_error})), $\big\{\mathcal{C}_{2,n}\big\}_{n\in\mathbb{N}}$ is also reliable when using it to transmit over the WTC I. Therefore, to establish \eqref{EQ:WTCI_UB_WTCII}, it suffices to show that for every $\epsilon>0$, there is an $n^\star\in\mathbb{N}$, such that for every $n>n^\star$
\begin{equation}
I_{\mathcal{C}_{2,n}}\big(M;\mathbf{Z}_1\big)\leq \epsilon,\label{EQ:WTCI_strong_secrecy_epsilon}
\end{equation}
where $\mathbf{Z}_1$ denoted the channel output sequences observed by the eavesdroppers of the WTC I. In other words, we show that the sequence of codes $\big\{\mathcal{C}_{2,n}\big\}_{n\in\mathbb{N}}$, designed to achieve strong-secrecy for the WTC II, also achieves strong-secrecy for the WTC I.

Let $\mathbf{Z}_2$ be the channel output observed by the eavesdroppers of the WTC II, fix $\epsilon>0$ and let $n_0\in\mathbb{N}$ be such that for every $n>n_0$,
\begin{equation}
\max_{\substack{\mathcal{S}\subseteq[1:n]:\\|\mathcal{S}|=\mu}}I_{\mathcal{C}_{2,n}}\big(M;\mathbf{Z}_2\big)\leq \frac{\epsilon}{2}.\label{EQ:WTCII_strong_secrecy_epsilon}
\end{equation}
For every $\mathbf{z}\in\mathcal{Z}^n$, where $\mathcal{Z}\triangleq\mathcal{X}\cup\{?\}$, define
\begin{equation}
\mathcal{A}(\mathbf{z})\triangleq\big\{i\in[1:n]\big|z_i=?\big\},
\end{equation}
and let $\Theta(\mathbf{Z})$ be
\begin{equation}
\Theta(\mathbf{Z})\triangleq\mathds{1}_{\big\{|\mathcal{A}(\mathbf{Z})|\leq\lceil \bar{\alpha}n\rceil\big\}}.\label{EQ:Theta_def}
\end{equation}
Namely, $\Theta$ indicates if the number of erasures in a sequence $\mathbf{z}\in\mathcal{Z}^n$ is greater than or equal to $\lceil\bar{\alpha}n\rceil$ or not.

By conditioning the mutual information term from \eqref{EQ:WTCI_strong_secrecy_epsilon} on $\Theta(\mathbf{Z}_1)$, we distinguish between the two cases of $\mathbf{Z}_1$ being better or worse than $\mathbf{Z}_2$ in terms of the number of erased symbols. When $\Theta(\mathbf{Z}_1)=0$, i.e., $\mathbf{Z}_1$ is worse that $\mathbf{Z}_2$, security for the WTC I is ensured since $\big\{\mathcal{C}_{2,n}\big\}_{n\in\mathbb{N}}$ achieve security for the WTC II. Otherwise, for the case that $\Theta(\mathbf{Z}_1)=1$, where $\mathbf{Z}_1$ is better than $\mathbf{Z}_2$, we use Sanov's Theorem to show that the probability of such an event exponentially decreases with the blocklength $n$, while the mutual information grows linearly at most. For any $n\in\mathbb{N}$, we have
\begin{align*}
I_{\mathcal{C}_{2,n}}\big(M;\mathbf{Z}_1\big)&\stackrel{(a)}=I_{\mathcal{C}_{2,n}}\Big(M;\Theta\big(\mathbf{Z}_1\big),\mathbf{Z}_1\Big)\\
&\stackrel{(b)}=I_{\mathcal{C}_{2,n}}\Big(M;\mathbf{Z}_1\Big|\Theta\big(\mathbf{Z}_1\big)\Big)\\
&=\underbrace{\mathbb{P}\Big(\Theta\big(\mathbf{Z}_1\big)\mspace{-2mu}=\mspace{-2mu}0\Big)I_{\mathcal{C}_{2,n}}\Big(M;\mathbf{Z}_1\Big|\Theta\big(\mathbf{Z}_1\big)\mspace{-2mu}=\mspace{-2mu}0\Big)}_{\mathcal{I}_0}\\
&+\underbrace{\mathbb{P}\Big(\Theta\big(\mathbf{Z}_1\big)\mspace{-2mu}=\mspace{-2mu}1\Big)I_{\mathcal{C}_{2,n}}\Big(M;\mathbf{Z}_1\Big|\Theta\big(\mathbf{Z}_1\big)\mspace{-2mu}=\mspace{-2mu}1\Big)}_{\mathcal{I}_1},\numberthis\label{EQ:WTCI_secrecY_theta}
\end{align*}
where (a) is because $\Theta\big(\mathbf{Z}_1\big)$ is a function of $\mathbf{Z}_1$, while (b) follows since the number of erasures in the output sequence of a DM-EC is defined by an i.i.d. process that is independent of the message.
%
%


For $\mathcal{I}_0$, taking any $n>n_0$, \eqref{EQ:WTCII_strong_secrecy_epsilon} implies that
\begin{equation}
I_{\mathcal{C}_{2,n}}\Big(M;\mathbf{Z}_1\Big|\Theta\big(\mathbf{Z}_1\big)=0\Big)\leq\max_{\substack{\mathcal{S}\subseteq[1:n]:\\|\mathcal{S}|=\mu}}I_{\mathcal{C}_{2,n}}\big(M;\mathbf{Z}_2\big)\leq \frac{\epsilon}{2}.\label{EQ:I1_UB}
\end{equation}
To upper bound $\mathcal{I}_1$, first note that
\begin{equation}
I_{\mathcal{C}_{2,n}}\Big(M;\mathbf{Z}_1\Big|\Theta\big(\mathbf{Z}_1\big)=1\Big)\leq n\log\big(|\mathcal{X}|+1\big),\label{EQ:I2_information_UB}
\end{equation}
holds for every $n\in\mathbb{N}$. Now, fix any  $\delta\in(\beta,\alpha)$; there exists an $n_1(\delta)\in\mathbb{N}$, such that for all $n>n_1$
\begin{equation}
    \lceil\bar{\alpha}n\rceil\leq\bar{\delta}n<\bar{\beta}n.\label{EQ:delta_in_between}
\end{equation}
Thus, for every $n>n_1(\delta)$ Sanov's Theorem \cite[Theorem 11.4.1]{Cover_Thomas} implies
\begin{equation}
\mathbb{P}\Big(\Theta\big(\mathbf{Z}_1\big)\mspace{-2mu}=\mspace{-2mu}1\Big)\leq\mathbb{P}\Big(\big|\mathcal{A}(\mathbf{Z}_1)\big|\leq\bar{\delta}n\Big)\leq(n+1)^2\cdot 2^{-nD_b(\delta,\beta)},\label{EQ:probability_UB_Sanov}
\end{equation}
where $D_b(\delta,\beta)=\alpha\log\left(\nicefrac{\delta}{\beta}\right)+\bar{\delta}\log\left(\nicefrac{\bar{\delta}}{\bar{\beta}}\right)$ is the relative entropy between the PMFs of two binary random variables distributed according to$\ber(\delta)$ and$\ber(\beta)$, respectively. Since $\delta\neq\beta$, we have that $D_b(\delta,\beta)>0$, and therefore, there is an $n_1(\delta)<n_2\in\mathbb{N}$, such that for every $n>n_2$,
\begin{equation}
\mathcal{I}_2\leq (n+1)^2\cdot 2^{-nD_b(\delta,\beta)}\cdot n\log\big(|\mathcal{X}|+1\big)\leq \frac{\epsilon}{2}.\label{EQ:probability_UB_Sanov_epsilon}
\end{equation}
%

Set $n^\star=\max\{n_0,n_2\}$. Based on (\ref{EQ:I1_UB}) and (\ref{EQ:probability_UB_Sanov_epsilon}), for every $n>n^\star$, we have
\begin{equation}
I_{\mathcal{C}_{2,n}}\big(M;\mathbf{Z}_1\big)=\mathcal{I}_0+\mathcal{I}_1\leq\epsilon,\label{EQ:WTCI_Secrecy_theta_UB}
\end{equation}
which completes the proof.



\subsubsection{Theorem \ref{TM:WTCII_capacity_WTC} - Converse}\label{SUBSUBSEC:WTCII_proof_theorem1_converse}

For the converse, we first show that with respect to the notations used in Proposition~\ref{PROP:WTCI_UB_WTCII},
\begin{equation}
C_\mathrm{S}^\mathrm{II}(\alpha)\leq C_\mathrm{S}^\mathrm{I}(\alpha)=\max_{\substack{Q_{U,X}:\\U-X-Y}}\Big[I(U;Y)-\alpha I(U;X)\Big],\label{EQ:WTCI_UB_WTCII_converse}
\end{equation}
for any $\alpha\in[0,1]$. For $\alpha=0,1$, the relation is straightforward as
\begin{subequations}
\begin{align}
C_\mathrm{S}^\mathrm{I}(0)&=\max_{Q_X}I(X;Y)=C_\mathrm{S}^\mathrm{II}(0)\\
C_\mathrm{S}^\mathrm{I}(1)&=0=C_\mathrm{S}^\mathrm{II}(1).
\end{align}
\end{subequations}
For $\alpha\in(0,1)$, \eqref{EQ:WTCI_UB_WTCII_converse} is established by relying on Proposition \ref{PROP:WTCI_UB_WTCII} and the continuity argument from Lemma \ref{LEMMA:continues}. Namely, by taking the limit of \eqref{EQ:WTCI_UB_WTCII} as $\beta\uparrow\alpha$ establishes \eqref{EQ:WTCI_UB_WTCII_converse}.

Having this, the converse follows by arguments similar to those presented in Section \ref{SUBSUBSEC:WTCI_proof_theorem1}. Fix $\alpha\in[0,1]$ and let $R\in\mathbb{R}_+$ be an $\alpha$-SS-achievable rate for the WTC II and $\big\{\mathcal{C}_n\big\}_{n\in\mathbb{N}}$ be its corresponding $(n,R)$ sequence of codes. By the definitions in (\ref{EQ:WTCI_error_prob}) and (\ref{EQ:WTCII_security_metric}), $\big\{\mathcal{C}_n\big\}_{n\in\mathbb{N}}$ are reliable and $\alpha$-semantically-secure for every message distribution, and in particular, for a uniform message distribution. This implies
\begin{equation}
C_{\mathrm{Sem}}(\alpha)\leq C_S^{\mathrm{II}}(\alpha)\leq \max_{\substack{Q_{U,X}:\\U-X-Y}}\Big[I(U;Y)-\alpha I(U;X)\Big]
\end{equation}
and completes the converse proof.

\begin{remark}
Our converse proof relies on the achievability being defined in terms of a limit as $n\to\infty$ (see Definition \ref{DEF:achievability_WTCII}). Namely, we show that in the limit, the eavesdropper in the WTC I setting is likely to be within a slightly higher channel-observation budget than this of the WTC II, which by continuity won't result in much extra rate. The chance of having too many channel observations is too small to provide non-negligible extra information. If, however, the blocklength $n$ can be chosen as a design parameter, then it may be possible that a finite $n$ results in a higher achievable secrecy-rate. For instance, notice that the optimal code of length $2n$ in not necessarily better than the optimal code of length $n$, since when the blocklenght is longer the eavesdropper has more flexibility in choosing his observations.
\end{remark}


\subsubsection{Theorem \ref{TM:WTCII_capacity_WTC} - Direct Part}\label{SUBSUBSEC:WTCII_proof_theorem1_direct}

As before, we start by showing the achievability of (\ref{EQ:WTCII_capacity_WTC}) when $U=X$. After doing so, we use channel prefixing to extend the proof to any $U$ with $U-X-Y$.

Fix $\alpha\in[0,1]$, $\epsilon>0$ and a PMF $Q_X$ on $\mathcal{X}$. Letting $M$ and $W$ be independent random variables uniformly distributed over $\mathcal{M}$ and $\mathcal{W}=[1:2^{n\tilde{R}}]$, respectively, we repeat the code construction from Section \ref{SUBSUBSEC:WTCI_proof_theorem1}. A similar analysis of the average error probability shows that if
\begin{equation}
R+\tilde{R}<I(X;Y),\label{EQ:WTCII_reliability_constraint}
\end{equation}
then
\begin{equation}
\mathbb{E}_{\mathbb{B}_n}\frac{1}{|\mathcal{M}|}\sum_{m\in\mathcal{M}}e_m(\mathbb{C}_n)\xrightarrow[n\to\infty]{}0,\label{EQ:WTCII_average_error_prob_vanish}
\end{equation}
where $\mathbb{C}_n$ is the random code that corresponds to the random codebook $\mathbb{B}_n$.


\par \textbf{Security Analysis:} Fix $\mathcal{S}\subseteq[1:n]$ with $|\mathcal{S}|=\mu=\lfloor \alpha n \rfloor$, recall that $\mathcal{Z}\triangleq\mathcal{X}\cup\{?\}$ and define the following PMF on $\mathcal{Z}^n$,
\begin{equation}
\Gamma^{(\mathcal{S})}_{\mathbf{Z}}(\mathbf{z})=\prod_{j\in\mathcal{S}^c}\mathds{1}_{\big\{z_j=?\big\}}\prod_{j\in\mathcal{S}}\mathcal{I}_Z(z_j),\quad\forall\mathbf{z}\in\mathcal{Z}^n,\label{EQ:WTCII_proof_gamma}
\end{equation}
where $\mathcal{I}_Z$ is the average output PMF of the identity DMC on $\mathcal{X}$, i.e.,
\begin{equation}
\mathcal{I}_Z(z)=\sum_{x\in\mathcal{X}}Q_X(x)\mathds{1}_{\{z=x\}}=\begin{cases}Q_X(z),\ z\in\mathcal{X}\\0,\ \ \ \ \ \ \ \mspace{5mu}z=?\end{cases}.
\end{equation}

For any $\mathcal{C}_n$ (defined by fixing $\mathcal{B}_n$) and $P_M\in\mathcal{P}(\mathcal{M})$, the relative entropy chain rule implies
\begin{align*}
I_{\mathcal{C}_n}(M;\mathbf{Z})&=D\left(P^{(\mathcal{C}_n,\mathcal{S})}_{\mathbf{Z}|M}\Big|\Big|P^{(\mathcal{C}_n,\mathcal{S})}_{\mathbf{Z}}\Big|P_M\right)\\
&=D\left(P^{(\mathcal{C}_n,\mathcal{S})}_{\mathbf{Z}|M}\Big|\Big|\Gamma^{(\mathcal{S})}_{\mathbf{Z}}\Big|P_M\right)\\
&\mspace{50mu}-D\left(P^{(\mathcal{C}_n,\mathcal{S})}_{\mathbf{Z}}\Big|\Big|\Gamma^{(\mathcal{S})}_{\mathbf{Z}}\Big|P_M\right),\numberthis\label{EQ:WTCII_divergence_gamma_grows}
\end{align*}
and therefore
\begin{align*}
\max_{P_M\in\mathcal{P}(\mathcal{M})}&I_{\mathcal{C}_n}(M;\mathbf{Z})\\
&\leq\max_{P_M\in\mathcal{P}(\mathcal{M})}D\left(P^{(\mathcal{C}_n,\mathcal{S})}_{\mathbf{Z}|M}\Big|\Big|\Gamma^{(\mathcal{S})}_{\mathbf{Z}}\Big|P_M\right)\\
&\leq\max_{P_M\in\mathcal{P}(\mathcal{M})}\sum_{m\in\mathcal{M}}P(m)\max_{\tilde{m}\in\mathcal{M}}D\left(P^{(\mathcal{C}_n,\mathcal{S})}_{\mathbf{Z}|M=\tilde{m}}\Big|\Big|\Gamma^{(\mathcal{S})}_{\mathbf{Z}}\right)\\
&=\max_{m\in\mathcal{M}}D\left(P^{(\mathcal{C}_n,\mathcal{S})}_{\mathbf{Z}|M=m}\Big|\Big|\Gamma^{(\mathcal{S})}_{\mathbf{Z}}\right).\numberthis\label{EQ:WTCII_divergence_maxm_grows}
\end{align*}

For any $\emptyset\neq\mathcal{A}\subseteq[1\mspace{-2mu}:\mspace{-2mu}n]$ and $\mathbf{z}\in\mathcal{Z}^n$, recall that $\mathbf{z}^\mathcal{A}\triangleq (z_i)_{i\in\mathcal{A}}$ is the sub-vector of $\mathbf{z}$ indexed by the elements of $\mathcal{A}$. The relative entropy chain rule further simplifies the RHS of \eqref{EQ:WTCII_divergence_maxm_grows} as follows. For any $m\in\mathcal{M}$, we have
\begin{align*}
D\left(P^{(\mathcal{C}_n,\mathcal{S})}_{\mathbf{Z}|M=m}\Big|\Big|\Gamma^{(\mathcal{S})}_{\mathbf{Z}}\right)&=D\left(P^{(\mathcal{C}_n,\mathcal{S})}_{\mathbf{Z}^\mathcal{S},\mathbf{Z}^{\mathcal{S}^c}|M=m}\Big|\Big|\Gamma^{(\mathcal{S})}_{\mathbf{Z}^\mathcal{S},\mathbf{Z}^{\mathcal{S}^c}}\right)\\
&=D\left(P^{(\mathcal{C}_n,\mathcal{S})}_{\mathbf{Z}^\mathcal{S}|M=m}\Big|\Big|\Gamma^{(\mathcal{S})}_{\mathbf{Z}^\mathcal{S}}\right)\\&+D\left(P^{(\mathcal{C}_n,\mathcal{S})}_{\mathbf{Z}^{\mathcal{S}^c}|M=m,\mathbf{Z}^\mathcal{S}}\Big|\Big|\Gamma^{(\mathcal{S})}_{\mathbf{Z}^{\mathcal{S}^c}}\Big|P^{(\mathcal{C}_n,\mathcal{S})}_{\mathbf{Z}^\mathcal{S}|M=m}\right)\\
&\stackrel{(a)}=D\left(P^{(\mathcal{C}_n,\mathcal{S})}_{\mathbf{Z}^\mathcal{S}|M=m}\Big|\Big|\Gamma^{(\mathcal{S})}_{\mathbf{Z}^\mathcal{S}}\right)\\
&\stackrel{(b)}=D\left(P^{(\mathcal{C}_n,\mathcal{S})}_{\mathbf{Z}^\mathcal{S}|M=m}\Big|\Big|\mathcal{I}^\mu_Z\right),\numberthis\label{EQ:WTCII_divergence_equality}
\end{align*}
where (a) is because $P^{(\mathcal{C}_n,\mathcal{S})}_{\mathbf{Z}^{\mathcal{S}^c}|M=m,\mathbf{Z}^\mathcal{S}=\mathbf{z}^\mathcal{S}}=\mathds{1}_{\big\{Z_i=?,\ i\in\mathcal{S}^c\big\}}=\Gamma^{(\mathcal{S})}_{\mathbf{Z}^{\mathcal{S}^c}}$, for every $\mathbf{z}^\mathcal{S}\in\mathcal{Z}^{|\mathcal{S}|}$, and (b) follows from \eqref{EQ:WTCII_proof_gamma}.


Combining \eqref{EQ:WTCII_divergence_gamma_grows}-\eqref{EQ:WTCII_divergence_equality}, we have that for every $\mathcal{C}_n$ and $\mathcal{S}\subseteq[1:n]$, with $|\mathcal{S}|=\mu=\lfloor \alpha n \rfloor$,
\begin{align*}
\max_{P_M\in\mathcal{P}(\mathcal{M})}D\Big(P^{(\mathcal{C}_n,\mathcal{S})}_{\mathbf{Z}|M}\Big|\Big|P^{(\mathcal{C}_n,\mathcal{S})}_{\mathbf{Z}}&\Big|P_M\Big)\\&\leq\max_{m\in\mathcal{M}}D\left(P^{(\mathcal{C}_n,\mathcal{S})}_{\mathbf{Z}^\mathcal{S}|M=m}\Big|\Big|\mathcal{I}^\mu_Z\right).\numberthis\label{EQ:WTCII_divergence_UB_nomaxS}
\end{align*}
In particular, \eqref{EQ:WTCII_divergence_UB_nomaxS} also holds when maximizing over the substes $S$, which gives
\begin{equation}
\mathrm{Sem}_{\mu}(\mathcal{C}_n)\leq \max_{\substack{m\in\mathcal{M},\\\mathcal{S}\subseteq[1:n]:\ |\mathcal{S}|=\mu}}D\left(P^{(\mathcal{C}_n,\mathcal{S})}_{\mathbf{Z}^\mathcal{S}|M=m}\Big|\Big|\mathcal{I}^\mu_Z\right).\label{EQ:WTCII_divergence_UB_maxmS}
\end{equation}

Having \eqref{EQ:WTCII_divergence_UB_maxmS}, let $\tilde{\delta}$ be an arbitrary positive real number to be determined later and consider the following probability.
\begin{align*}
&\mathbb{P}\bigg(\Big\{\mathrm{Sem}_{\mu}(\mathbb{C}_n)\leq e^{-n\tilde{\delta}}\Big\}^c\bigg)\\
&=\mathbb{P}\Bigg(\max_{\substack{P_M\in\mathcal{P}(\mathcal{M}),\\\mathcal{S}\subseteq[1:n]:\ |\mathcal{S}|=\mu}}D\left(P^{(\mathbb{C}_n,\mathcal{S})}_{\mathbf{Z}|M}\Big|\Big|P^{(\mathbb{C}_n,\mathcal{S})}_{\mathbf{Z}}\Big|P_M\right)>e^{-n\tilde{\delta}}\Bigg)\\
&\stackrel{(a)}\leq\mathbb{P}\left(\max_{\substack{m\in\mathcal{M},\\\mathcal{S}\subseteq[1:n]:\ |\mathcal{S}|=\mu}}D\left(P^{(\mathbb{C}_n,\mathcal{S})}_{\mathbf{Z}^\mathcal{S}|M=m}\Big|\Big|\mathcal{I}^\mu_Z\right)>e^{-n\tilde{\delta}}\right)\\
&=\mathbb{P}\left(\bigcup_{\substack{m\in\mathcal{M},\\\mathcal{S}\subseteq[1:n]:\ |\mathcal{S}|=\mu}}\bigg\{D\left(P^{(\mathbb{C}_n,\mathcal{S})}_{\mathbf{Z}^\mathcal{S}|M=m}\Big|\Big|\mathcal{I}^\mu_Z\right)>e^{-n\tilde{\delta}}\bigg\}\right)\\
&\stackrel{(b)}\leq\sum_{\substack{m\in\mathcal{M},\\\mathcal{S}\subseteq[1:n]:\ |\mathcal{S}|=\mu}}\mathbb{P}\bigg(D\left(P^{(\mathbb{C}_n,\mathcal{S})}_{\mathbf{Z}^\mathcal{S}|M=m}\Big|\Big|\mathcal{I}^\mu_Z\right)> e^{-n\tilde{\delta}}\bigg),\numberthis\label{EQ:WTCII_final_UB}
\end{align*}
where (a) uses \eqref{EQ:WTCII_divergence_UB_maxmS}, and (b) is the union bound.

Each term in the sum on the RHS of \eqref{EQ:WTCII_final_UB} falls into the framework of the stronger soft-covering lemma, with respect to a blocklength of $\mu$ and the identity channel. Noting that $|\mathcal{W}|=2^{n\tilde{R}}=2^{\mu\frac{n\tilde{R}}{\mu}}$, we have that as long as
\begin{equation}
\frac{n\tilde{R}}{\mu}>H(X)\label{EQ:WTCII_secrecy_constraint},
\end{equation}
there exist $\delta_1,\delta_2>0$ that for sufficiently large $n$ satisfy
\begin{equation}
\mathbb{P}\bigg(D\left(P^{(\mathbb{C}_n,\mathcal{S})}_{\mathbf{Z}^\mathcal{S}|M=m}\Big|\Big|\mathcal{I}^\mu_Z\right)> e^{-n\delta_1}\bigg)\leq e^{-e^{n\delta_2}}.\label{EQ:WTCII_secrecy_soft_covering_result}
\end{equation}
Since $\mu=\lfloor \alpha n \rfloor\leq\alpha n$, taking
\begin{equation}
\tilde{R}>\alpha H(X)\label{EQ:WTCII_secrecy_constraint3},
\end{equation}
is sufficient to satisfy \eqref{EQ:WTCII_secrecy_constraint} for every $n\in\mathbb{N}$. 

Setting $\tilde{\delta}=\delta_1$ and plugging \eqref{EQ:WTCII_secrecy_soft_covering_result} into \eqref{EQ:WTCII_final_UB}, gives
\begin{align*}
\mathbb{P}\bigg(\Big\{\mathrm{Sem}_{\mu}(\mathbb{C}_n)\leq e^{-n\delta_1}\Big\}^c\bigg)&\leq \sum_{\substack{m\in\mathcal{M},\\\mathcal{S}\subseteq[1:n]:\ |\mathcal{S}|=\mu}}e^{-e^{n\delta_2}}\\
&\leq2^n\cdot2^{nR}\cdot e^{-e^{n\delta_2}}\\
&\triangleq \kappa_n\xrightarrow[n\to\infty]{}0.\label{EQ:WTCII_secrecy_comp_probability_final}
\end{align*}

Invoking Lemma \ref{LEMMA:WTCI_selection} once more, we have that if (\ref{EQ:WTCII_reliability_constraint}) and (\ref{EQ:WTCII_secrecy_constraint}) are satisfied, then there is a sequence of $(n,R)$ $\alpha$-semantically-secure codes $\big\{\mathcal{C}_n\big\}_{n\in\mathbb{N}}$, with
\begin{equation}
\mathbb{E}_{\mathbb{C}_n}\frac{1}{|\mathcal{M}|}\sum_{m\in\mathcal{M}}e_m(\mathbb{C}_n)\xrightarrow[n\to\infty]{}0.\label{EQ:WTCII_average_error_prob_vanish_fixed_code}
\end{equation}
The pruning argument from Section \ref{SUBSUBSEC:WTCI_proof_theorem1} again upgrades $\big\{\mathcal{C}_n\big\}_{n\in\mathbb{N}}$ to be reliable with respect to the maximal error probability. Combining \eqref{EQ:WTCII_reliability_constraint} and \eqref{EQ:WTCII_secrecy_constraint} shows the achievability of
\begin{equation}
R<\max_{Q_X}\Big[I(X;Y)-\alpha H(X)\Big].
\end{equation}

Finally, we prefix a DMC $Q_{X|U}$ to the original WTC II to obtain a new main channel $Q_{Y|U}$, given by
\begin{equation}
Q^n_{Y|U}(\mathbf{y}|\mathbf{u})=\sum_{\mathbf{x}\in\mathcal{X}^n}Q^n_{X|U}(\mathbf{x}|\mathbf{u})Q^n_{Y|X}(\mathbf{y}|\mathbf{x}).
\end{equation}
Furthermore, $\Gamma_\mathbf{Z}^{(\mathcal{S})}$ from (\ref{EQ:WTCII_proof_gamma}) is redefined as
\begin{equation}
\Gamma^{(\mathcal{S})}_{\mathbf{Z}}(\mathbf{z})=\prod_{j\in\mathcal{S}^c}\mathds{1}_{\big\{z_j=?\big\}}\prod_{j\in\mathcal{S}}Q_Z(z_j),\ \ \forall\mathbf{z}\in\mathcal{Z}^n,\label{EQ:WTCII_proof_gamma_new}
\end{equation}
where $Q_Z$ is given by
\begin{align*}
Q_Z(z)&=\sum_{(u,x)\in\mathcal{U}\times\mathcal{X}}Q_U(u)Q_{X|U}(x|u)\mathds{1}_{\{z=x\}}\\
&=\begin{cases}\sum_{u\in\mathcal{V}}Q_U(u)Q_{X|U}(z|u),\ z\in\mathcal{X}\\0,\ \ \ \ \ \ \ \ \ \ \ \ \ \ \ \ \ \ \ \ \ \ \ \ \ \ \ \ \mspace{5mu}z=?\end{cases}.
\end{align*}
Repeating a similar analysis as above shows that reliability is achieved if
\begin{equation}
R+\tilde{R}<I(U;Y),\label{EQ:WTCII_prefix_RB1}
\end{equation}
while the rate needed for the stronger soft-covering lemma is
\begin{equation}
\tilde{R}>\alpha I(U;X).\label{EQ:WTCII_prefix_RB2}
\end{equation}
Putting \eqref{EQ:WTCII_prefix_RB1}-\eqref{EQ:WTCII_prefix_RB2} together yields that any rate $R\in\mathbb{R}_+$ satisfying
\begin{equation}
R<\max_{\substack{Q_{U,X}:\\U-X-Y}}\Big[I(U;Y)-\alpha I(U;X)\Big]\label{EQ:WTCII_achievable_rate_V},
\end{equation}
is strongly $\alpha$-SS-achievable and concludes the proof.


\section{Summary and Concluding Remarks}\label{SEC:summary}

We derived the SS capacity of the WTC II with a noisy main channel. The SS metric ensures that the unnormalized mutual information between the message and the eavesdropper's observation is arbitrarily small, even when maximized over all message distributions and all possible choices of the eavesdropper's observation. The main tool used in the direct proof is a novel and stronger version of Wyner's soft covering lemma, that states that a random codebook achieves the soft-covering phenomenon with high probability as long as its rate is higher than the mutual information between the input and output of the DMC. Furthermore, the probability of failure is doubly-exponentially small in the blocklength, thus making the lemma advantageous in proving the existence of codebooks that satisfy exponentially many constraints. A code that achieves SS for the considered WTC II should do just that.

The SS capacity was achieved by using classic Wyner's wiretap codes. Since the combined number of messages and subsets grows only exponentially with the blocklength, SS was established by applying the union bound and invoking the stronger soft-covering lemma. The direct proof showed that rates up to the weak-secrecy capacity of the WTC I with a DM-EC to the eavesdropper are achievable. The converse followed by showing that the capacity of this WTC I is an upper bound on the SS capacity of the WTC II.

As a preliminary and simple application of the stronger soft-covering lemma, it was used to achieve SS for the WTC I. A main goal in doing so was to emphasize the advantage of this approach over other methods for achieving SS for this scenario, such as the expurgation technique. While the expurgation method fails to generalize to some multiuser settings, such as the multiple access WTC, an achievability proof that relies on the stronger soft-covering lemma goes through by similar steps to those presented here. Thus making the stronger soft-covering lemma a tool by which the common weak-secrecy and strong-secrecy results can be upgraded to SS. Furthermore, the lemma might prove useful in any other scenario in which performance is measures with respect to an exponential number of constraints.


\section*{Acknowledgment}

The authors would like to thank Mohamed Nafea and Aylin Yener for a helpful discussion of the problem.



\appendices


\section{Proof of Lemma \ref{LEMMA:soft_covering_stronger}}\label{APPEN:soft_covering_stronger_proof}

Let $n_0\in\mathbb{N}$ be such that \eqref{EQ:soft_covering} holds for any $n>n_0$. For these values of $n$ we have
\begin{align*}
&\mathbb{E}_{\mathbb{B}_n}D\Big(P^{(\mathbb{B}_n)}_{\mathbf{V}}\Big|\Big|Q_V^n\Big)\\
&\begin{multlined}[b][.9\columnwidth]=\mathbb{E}_{\mathbb{C}_n}\Bigg[D\Big(P^{(\mathbb{B}_n)}_{\mathbf{V}}\Big|\Big|Q_V^n\Big)\bigg(\mathds{1}_{\big\{D\big(P^{(\mathbb{B}_n)}_{\mathbf{V}}\big|\big|Q_V^n\big)\leq e^{-n\gamma_1}\big\}}\\+\mathds{1}_{\big\{D\big(P^{(\mathbb{B}_n)}_{\mathbf{V}}\big|\big|Q_V^n\big)>e^{-n\gamma_1}\big\}}\bigg)\Bigg]\end{multlined}\\
&\stackrel{(a)}\leq e^{-n\gamma_1}+n\log\left(\frac{1}{\mu_V}\right)\mathbb{P}\Big(D\Big(P^{(\mathbb{B}_n)}_{\mathbf{V}}\Big|\Big|Q_V^n\Big)>e^{-n\gamma_1}\Big)\\
&\stackrel{(b)}\leq e^{-n\gamma_1}+n\log\left(\frac{1}{\mu_V}\right)e^{-e^{n\gamma_2}},\numberthis\label{EQ:soft_covering_stronger_expectationUB}
\end{align*}
where (a) follows because for every fixed $\mathcal{B}_n$
\begin{align*}
D\Big(P^{(\mathbb{B}_n)}_{\mathbf{V}}\Big|\Big|Q_V^n\Big)&=\sum_{\mathbf{z}\in\mathcal{Z}^n}P^{(\mathcal{B}_n)}(\mathbf{z})\log\left(\frac{P^{(\mathcal{B}_n)}(\mathbf{v})}{Q_V^n(\mathbf{v})}\right)\\
&\leq n\log\left(\frac{1}{\mu_V}\right),
\end{align*}
and $\mu_v=\min_{v\in\supp(Q_V)}Q_V(v)$, while (b) follows from \eqref{EQ:soft_covering}.


\section{Proof of Lemma \ref{LEMMA:soft_covering_UB}}\label{APPEN:soft_covering_UB_proof}

Fix a codebook $\mathcal{C}_n$ and define
\begin{equation}
\Theta=\mathds{1}_{\big\{\big(\mathbf{U}(W,\mathcal{B}_n),\mathbf{V}\big)\notin\mathcal{A}_\epsilon\big\}}+1.\label{EQ:Theta_indicator_def}
\end{equation}
Note that for $\theta=1,2$, we have
\begin{align*}
P^{(\mathcal{B}_n)}_{\Theta}(\theta)&=\int dP^{(\mathcal{B}_n)}_{\mathbf{V}}\sum_{w\in\mathcal{W}}2^{-nR}Q_{V|U=\mathbf{u}(w,\mathcal{B}_n)}^n\\
&\begin{multlined}[b][.77\columnwidth]\times\bigg[\mathds{1}_{\big\{\theta=1\big\}\cap\big\{\big(\mathbf{u}(w,\mathcal{B}_n),\mathbf{V}\big)\in\mathcal{A}_\epsilon\big\}}\\+\mathds{1}_{\big\{\theta=2\big\}\cap\big\{\big(\mathbf{u}(w,\mathcal{B}_n),\mathbf{V}\big)\notin\mathcal{A}_\epsilon\big\}}\bigg]\end{multlined}\\
&=\int dP_{\mathcal{B}_n,\theta},\numberthis\label{EQ:P_theta}
\end{align*}
and consequently, for every measurable $\mathcal{A}\subseteq\mathcal{V}^n$,
\begin{align*}
\mathbb{P}_{P^{(\mathcal{B}_n)}_{\mathbf{V},\Theta}}\Big(\mathbf{V}\in\mathcal{A},\Theta=\theta\Big)&=
\mathbb{P}_{P_{\mathcal{B}_n,\theta}}\Big(\mathbf{V}\in\mathcal{A}\Big)\\
&=\int_\mathcal{A} dP_{\mathcal{B}_n,\theta}.\numberthis\label{EQ:P_theta_v}
\end{align*}

For simplicity of notation, denote $P^{(\mathcal{B}_n)}_{\mathbf{V}}\triangleq P$, $P_{\mathcal{B}_n,1}\triangleq P_1$, $P_{\mathcal{B}_n,2}\triangleq P_2$, $Q_V^n\triangleq Q$ and $P^{(\mathcal{B}_n)}_{\Theta}\triangleq \Gamma_{\Theta}$, and consider
\begin{align*}
D(P||Q)&=\int dP \log\left(\frac{dP}{dQ}\right)\\
       &\stackrel{(a)}=\int dQ \frac{dP}{dQ}\log\left(\frac{dP}{dQ}\right)\\
       &\begin{multlined}[b][.83\columnwidth]\stackrel{(b)}=\int dQ \mspace{3mu}\mathbb{E}_{\Gamma_{\Theta}}\mspace{-5mu}\left[\frac{1}{\Gamma_{\Theta}(\Theta)}\cdot\frac{dP_\Theta}{dQ}\right]\\\times\log\left(\mathbb{E}_{\Gamma_{\Theta}}\mspace{-5mu}\left[\frac{1}{\Gamma_{\Theta}(\Theta)}\cdot\frac{dP_\Theta}{dQ}\right]\right)\end{multlined}\\
       &\begin{multlined}[b][.83\columnwidth]\stackrel{(c)}\leq\int dQ \mspace{3mu}\mathbb{E}_{\Gamma_{\Theta}}\mspace{-5mu}\bigg[\frac{1}{\Gamma_{\Theta}(\Theta)}\cdot\frac{dP_\Theta}{dQ}\cdot\\\times\log\left(\frac{1}{\Gamma_{\Theta}(\Theta)}\cdot\frac{dP_\Theta}{dQ}\right)\bigg]\end{multlined}\\
       &\begin{multlined}[b][.83\columnwidth]=\sum_{\theta=1,2}\Gamma_{\Theta}(\theta)\int dQ \mspace{3mu}\frac{1}{\Gamma_{\Theta}(\theta)}\cdot\frac{dP_\theta}{dQ}\\\times\log\left(\frac{1}{\Gamma_{\Theta}(\theta)}\cdot\frac{dP_\theta}{dQ}\right)\end{multlined}\\
       &\begin{multlined}[b][.83\columnwidth]\stackrel{(d)}=\sum_{\theta=1,2}\log\left(\frac{1}{\Gamma_{\Theta}(\theta)}\right)\int dP_\theta\\+\sum_{\theta=1,2}\int dP_\theta \log\left(\frac{dP_\theta}{dQ}\right)\end{multlined}\\
       &\stackrel{(e)}=h\left(\int dP_1\right)+\sum_{\theta=1,2}\int dP_\theta \log\Delta_{\mathcal{B}_n,\theta},\numberthis
\end{align*}
where:\\
(a) follows since for any two measures $\mu,\lambda$ with $\mu\ll\lambda$ and a $\mu-$integrable function $g$, we have
 \begin{equation}
 \int g d\mu = \int g\frac{d\mu}{d\lambda} d\lambda;\label{EQ:RD_derivative_property}
 \end{equation}
(b) follows from \eqref{EQ:P_theta_v} and the law of total probability;\\
(c) follows by applying Jensen’s inequality to the convex function $x\mapsto x\log(x)$;\\
(d) follows by the properties of the logarithm and \eqref{EQ:RD_derivative_property};\\
(e) follows from \eqref{EQ:P_theta} and the definition of $\Delta_{\mathcal{B}_n,\theta}$, for $\theta=1,2$, in \eqref{EQ:delta_def}.


\section{Proof of the Chenoff Bound - Lemma \ref{LEMMA:Chernoff}}\label{APPEN:chernoff_proof}

Let $X$ have the same distribution as $X_1$. For any $\lambda>0$, we have

\begin{align*}
    \mathbb{P} \left( \frac{1}{M} \sum_{m=1}^M X_m \geq c \right) &\stackrel{(a)}{\leq} \frac{\mathbb{E} e^{\lambda \sum_{m=1}^M X_m}}{e^{\lambda cM}} \\
    &= \left( \frac{ \mathbb{E} e^{\lambda X}}{e^{\lambda c}}  \right)^M \\
    &\stackrel{(b)}{\leq} \left( \frac{ 1 + \frac{e^{\lambda B}-1}{B} \mathbb{E} X}{e^{\lambda c}}  \right)^M \\
    &\leq \left( \frac{ 1 + \frac{e^{\lambda B}-1}{B} \mu}{e^{\lambda c}}  \right)^M \\
    &\stackrel{(c)}{\leq} \left( \frac{ e^{\frac{e^{\lambda B}-1}{B} \mu}}{e^{\lambda c}}  \right)^M \\
    &= e^{-M \left( \lambda c + \frac{\mu}{B} - \frac{\mu e^{\lambda B}}{B} \right)},\numberthis\label{EQ:chernoff_proof_UB_lambda}
\end{align*}
where (a) is the Chernoff bound, (b) uses the fact that $e^{\lambda x} \leq 1 + \frac{e^{\lambda B}-1}{B}x$, for $x \in [0,B]$ due to convexity, and (c) uses $1 + x \leq e^x$.

Optimizing the RHS of \eqref{EQ:chernoff_proof_UB_lambda} over $\lambda$ given $\lambda^\star=\frac{1}{B} \ln \frac{c}{\mu}$ as the minimizer, as long as $\frac{c}{\mu} \geq 1$.  Plugging this into \eqref{EQ:chernoff_proof_UB_lambda} yields
\begin{equation}
\label{chernoff constant}
    \mathbb{P} \left( \frac{1}{M} \sum_{m=1}^M X_m \geq c \right) \leq e^{-\frac{M \mu}{B} \Big( \frac{c}{\mu} \left( \ln \frac{c}{\mu} - 1 \right) + 1 \Big)}, \quad \forall \frac{c}{\mu} \geq 1.
\end{equation}
This is a good bound when $\mu \ll B$, as it is in our case. If $c/\mu$ is shrinking, then to further simplify the bound consider the third order Tayler expansion of $x (\ln x - 1)$ about $x=1$,
\begin{equation}
    x (\ln x - 1) + 1 \geq \frac{1}{2} (x-1)^2 - \frac{1}{6} (x-1)^3, \quad \forall x \geq 1.\label{EQ:Taylor_expantion_x}
\end{equation}
The LHS in \eqref{EQ:Taylor_expantion_x} is a lower bound because the fourth derivative is positive for all $x \geq 1$. Furthermore, if $x-1 \leq 1$, we have
\begin{equation}
    \frac{1}{2} (x-1)^2 - \frac{1}{6} (x-1)^3 \geq \frac{1}{3} (x-1)^2,\quad \forall x \in [1,2].\label{EQ:Taylor_expantion_x2}
\end{equation}
Putting it all together gives
\begin{equation}
\label{chernoff shrinking}
    \mathbb{P} \left( \frac{1}{M} \sum_{m=1}^M X_m \geq c \right) \leq e^{-\frac{M \mu}{3 B} \left( \frac{c}{\mu} - 1 \right)^2}, \quad  \forall \frac{c}{\mu} \in [1,2].
\end{equation}


\section{Proof of Lemma \ref{LEMMA:WTCI_selection}}\label{APPEN:selection_proof}

Since $\big\{f_n^{(i)}\big\}_{n\in\mathbb{N}}$, $i\in[1:I]$, are bounded and by (\ref{EQ:WTCI_selection_it}), there exists a sequence $\{\delta_n\}_{n\in\mathbb{N}}$ such that
\begin{equation}
\mathbb{E}f_n^{(i)}(A_n)\leq \delta_n,\quad \forall i\in[1:I],\quad n\in\mathbb{N},
\end{equation}
and $\delta_n\to 0$ as $n\to\infty$. We have
\begin{align*}
\mathbb{P}\bigg(\bigcup_{i=1}^I\Big\{f^{(i)}_n(A_n)\geq(I+&1)\delta_n\Big\}\bigg)\\
&\leq \sum_{i=1}^I\mathbb{P}\Big(f^{(i)}_n(A_n)\geq(I+1)\delta_n\Big)\\
&\stackrel{(a)}\leq\sum_{i=1}^I\frac{\mathbb{E}f^{(i)}_n(A_n)}{(I+1)\delta_n}\\
&\stackrel{(b)}\leq\frac{I}{(I+1)}\\
&<1.\numberthis
\end{align*}
Therefore, there exists a realization $\{a_n\}_{n\in\mathbb{N}}$ of $\big\{A_n\big\}_{n\in\mathbb{N}}$ such that
\begin{equation}
f_n^{(i)}(a_n)<(I+1)\delta_n\triangleq\tilde{\delta}_n,\quad \forall i\in[1:I],\quad n\in\mathbb{N}.
\end{equation}
Since $I<\infty$ independently of $n$, we have $\tilde{\delta}_n\to 0$ as $n\to\infty$.


\section{Proof of Lemma \ref{LEMMA:continues}}\label{APPEN:contineous_proof}

We prove the continuity of $C_\mathrm{S}^\mathrm{I}(\beta)$ inside $(0,1)$ by showing that it is bounded and convex. Let $\beta_1,\beta_2\in(0,1)$, $\lambda\in[0,1]$ and observe that
\begin{align*}
C_\mathrm{S}^\mathrm{I}&(\lambda\beta_1+\bar{\lambda}\beta_2)\\
    &=\max_{\substack{Q_{U,X}:\\U-X-Y}}\Big[(\lambda+\bar{\lambda})I(U;Y)-(\lambda\beta_1+\bar{\lambda}\beta_2) I(U;X)\Big]\\
    &\leq\lambda\max_{\substack{Q_{U,X}:\\U-X-Y}}\Big[I(U;Y)-\beta_1I(U;X)\Big]\\
    &\mspace{140mu}+\bar{\lambda}\max_{\substack{Q_{U,X}:\\U-X-Y}}\Big[I(U;Y)-\beta_2I(U;X)\Big]\\
    &=\lambda C_\mathrm{S}^\mathrm{I}(\beta_1)+\bar{\lambda} C_\mathrm{S}^\mathrm{I}(\beta_2).\numberthis
\end{align*}
Furthermore, for every $\beta\in(0,1)$,
\begin{equation}
C_\mathrm{S}^\mathrm{I}(\beta)\leq\max_{Q_{X}} I(X;Y)\leq \log|\mathcal{Y}|<\infty.
\end{equation}

\newpage
\bibliographystyle{unsrt}
\bibliographystyle{IEEEtran}
\bibliography{ref}

\begin{IEEEbiographynophoto}{Ziv Goldfeld}
(S'13) received his B.Sc.\@ (summa cum laude) and M.Sc.\@ (summa cum laude) degrees in Electrical and Computer Engineering from the Ben-Gurion University, Israel, in 2012 and 2014, respectively. He is currently a student in the direct Ph.D. program for honor students in Electrical and Computer Engineering at that same institution.

Between 2003 and 2006, he served in the intelligence corps of the Israeli Defense Forces.

Ziv is a recipient of several awards, among them the Dean's List Award, the Basor Fellowship for honor students in the direct Ph.D. program, the Lev-Zion fellowship and the Minerva Short-Term Research Grant (MRG).
\end{IEEEbiographynophoto}

\begin{IEEEbiographynophoto}{Paul Cuff}
(S'08-M'10) received the B.S. degree in electrical engineering from Brigham Young University, Provo, UT, in 2004 and the M.S. and Ph. D. degrees in electrical engineering from Stanford University in 2006 and 2009. Since 2009 he has been an Assistant Professor of Electrical Engineering at Princeton University.

As a graduate student, Dr. Cuff was awarded the ISIT 2008 Student Paper Award for his work titled “Communication Requirements for Generating Correlated Random Variables” and was a recipient of the National Defense Science and Engineering Graduate Fellowship and the Numerical Technologies Fellowship. As faculty, he received the NSF Career Award in 2014 and the AFOSR Young Investigator Program Award in 2015.
\end{IEEEbiographynophoto}

\begin{IEEEbiographynophoto}{Haim H. Permuter}
(M'08-SM'13) received his B.Sc.\@ (summa cum laude) and M.Sc.\@ (summa cum laude) degrees in Electrical and Computer Engineering from the Ben-Gurion University, Israel, in 1997 and 2003, respectively, and the Ph.D. degree in Electrical Engineering from Stanford University, California in 2008.

Between 1997 and 2004, he was an officer at a research and development unit of the Israeli Defense Forces. Since 2009 he is with the department of Electrical and Computer Engineering at Ben-Gurion University where he is currently an associate professor.

Prof. Permuter is a recipient of several awards, among them the Fullbright Fellowship, the Stanford Graduate Fellowship (SGF), Allon Fellowship, and and the U.S.-Israel Binational Science Foundation Bergmann Memorial Award. Haim is currently serving on the editorial board of the IEEE Transactions on Information Theory.
\end{IEEEbiographynophoto}

\end{document}